\documentclass[useAMS,usenatbib]{mn2e}

\usepackage{graphicx,amssymb,color}
\usepackage[normalem]{ulem}

\title[Fates of Solar systems with wide planet]
{The fates of Solar system analogues with one additional distant planet}

\author[Veras]{
Dimitri Veras$^{1}$\thanks{E-mail: d.veras@warwick.ac.uk}
\\
$^{1}$Department of Physics, University of Warwick, Coventry CV4 7AL, UK
}

\pubyear{2016}

\begin{document}
\label{firstpage}
\pagerange{\pageref{firstpage}--\pageref{lastpage}}
\maketitle

\begin{abstract}
The potential existence of a distant planet (``Planet Nine'') in the Solar system  has prompted a re-think about the evolution of planetary systems. As the Sun transitions from a main sequence star into a white dwarf, Jupiter, Saturn, Uranus and Neptune are currently assumed to survive in expanded but otherwise unchanged orbits. However, a sufficiently-distant and sufficiently-massive extra planet would alter this quiescent end scenario through the combined effects of Solar giant branch mass loss and Galactic tides.  Here, I estimate bounds for the mass and orbit of a distant extra planet that would incite future instability in systems with a Sun-like star and giant planets with masses and orbits equivalent to those of Jupiter, Saturn, Uranus and Neptune. I find that this boundary is diffuse and strongly dependent on each of the distant planet's orbital parameters. Nevertheless, I claim that instability occurs more often than not when the planet is as massive as Jupiter and harbours a semimajor axis exceeding about 300 au, or has a mass of a super-Earth and a semimajor axis exceeding about 3000 au. These results hold for orbital pericentres ranging from 100 to at least 400 au. This instability scenario might represent a common occurrence, as potentially evidenced by the ubiquity of metal pollution in white dwarf atmospheres throughout the Galaxy.
\end{abstract}

\begin{keywords}
methods: numerical -- celestial mechanics --  planets and satellites: dynamical evolution and stability -- Sun: evolution -- stars: AGB and post-AGB -- stars: white dwarfs 
\end{keywords}

\section{Introduction}

Thousands of planets outside of our Solar system have already been discovered despite our potential ignorance of what planets may reside in our own backyard. This arresting notion \citep{broetal2004for3,glacha2006,iorio2012} was given added credence with the discovery of the second Sednoid, 2012 VP113 \citep{trushe2014}, because both that object and Sedna have arguments of pericentre clustered around $-50^{\circ}$ (see their table 1). This clustering is not likely to be a result of observational bias \citep{deldel2014}, and may be explained by one or more planets which exist beyond the orbit of Neptune but are so far invisible to us \citep{iorio2014,luhman2014,trushe2014,gometal2015for3,iorio2015}. However, the existence of these two scattered disc objects do not necessarily require the presence of additional planets \citep{jiletal2015}.

The recent finding that several distant Kuiper Belt Objects exhibit this same clustering in orbital space, as well a clustering in physical space \citep{batbro2016}, has prompted a febrile response amongst the public and scientific community. Constraining the location and size of a putative ``Planet Nine" has been the focus of many subsequent publications, through both orbital dynamics \citep{beust2016,brobat2016,deldel2016a,deldel2016b,deletal2016for3,fieetal2016,holpay2016a,holpay2016b,lawetal2016,maletal2016for3} and intrinsic physical properties \citep{cowetal2016for3,foretal2016,ginetal2016for3,linmor2016,toth2016}. 

Overall and roughly, these studies suggest that Planet Nine is more massive than the Earth and resides at hundreds or thousands of au away from the Sun (other similarly massive planets could lie further away). Other studies investigated potential origins for Planet Nine, including in-situ formation \citep{kenbro2016}, scattering into its current (theorized) orbit \citep{brokey2016} and capture from other stars in the Sun's birth cluster \citep{musetal2016for3}. The survivability of Planet Nine due to passing stars has also been investigated \citep{liada2016}. Additional planets which orbit the Sun also could exist but evidence for their presence has yet to be marshalled.

Two points not emphasized in the above studies are: (1) how our knowledge of the fate of the Solar system changes with the presence of planets beyond Neptune's orbit, and (2) the deeper fundamental issue of what the implications are for similarly-constructed systems and more generally exoplanetary science. These two concepts are linked through white dwarf planetary systems, and represent the motivation for this paper.

\subsection{The fate of an eight-planet Solar system}

\begin{figure*}
\centerline{
\includegraphics[width=8cm]{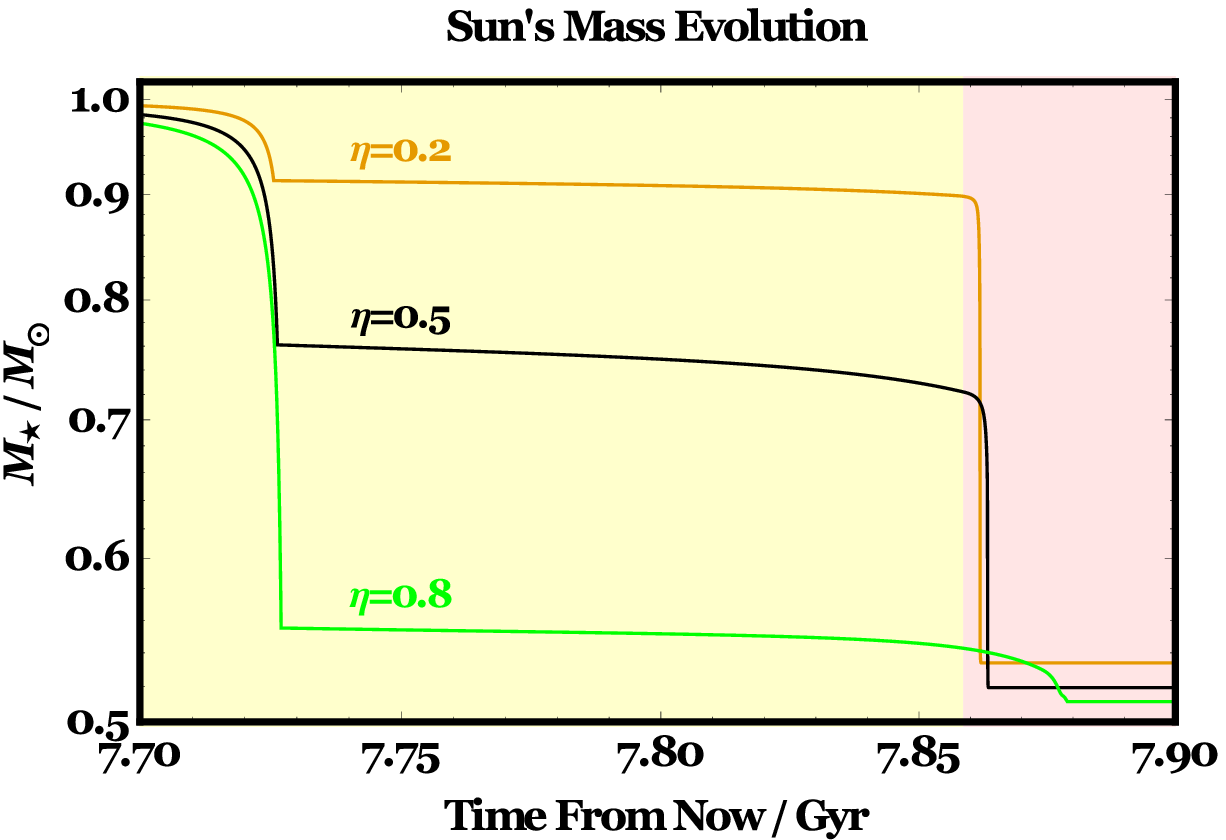}
\includegraphics[width=8cm]{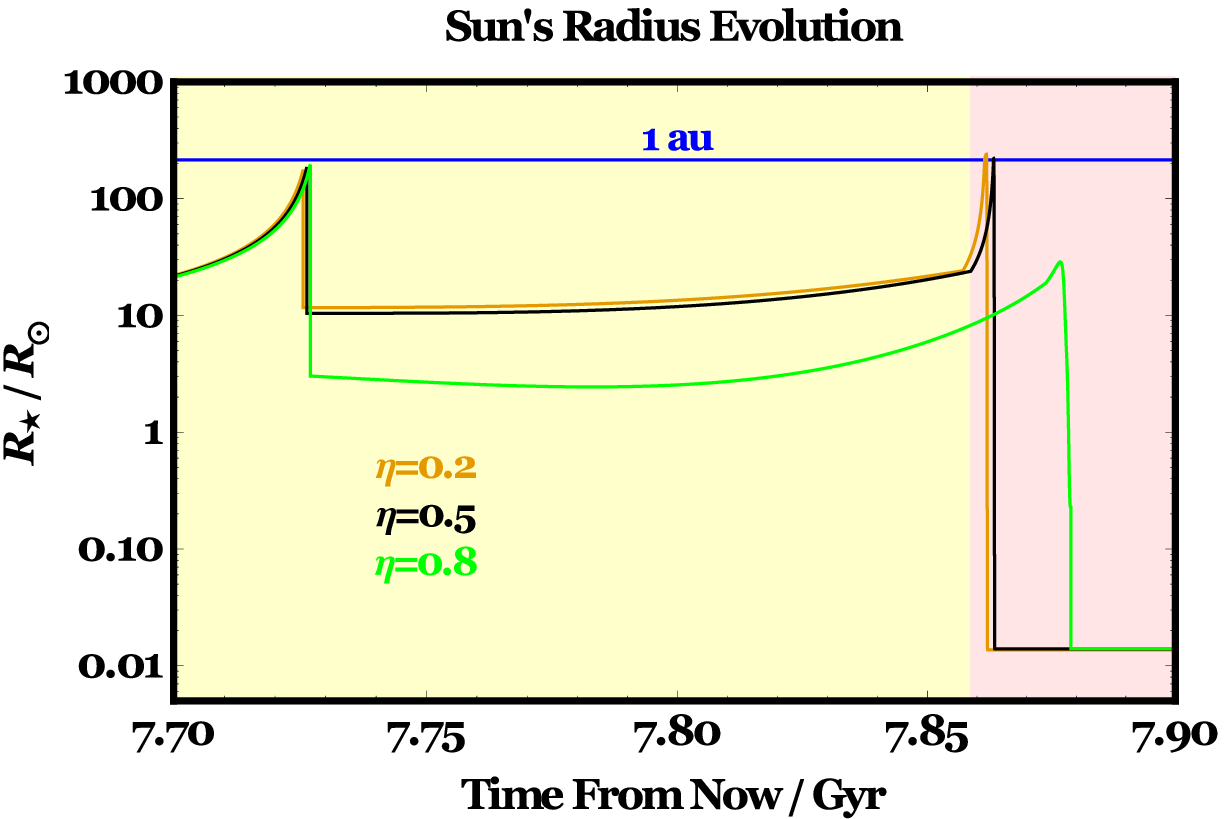}
}
\
\
\centerline{
\includegraphics[width=8cm]{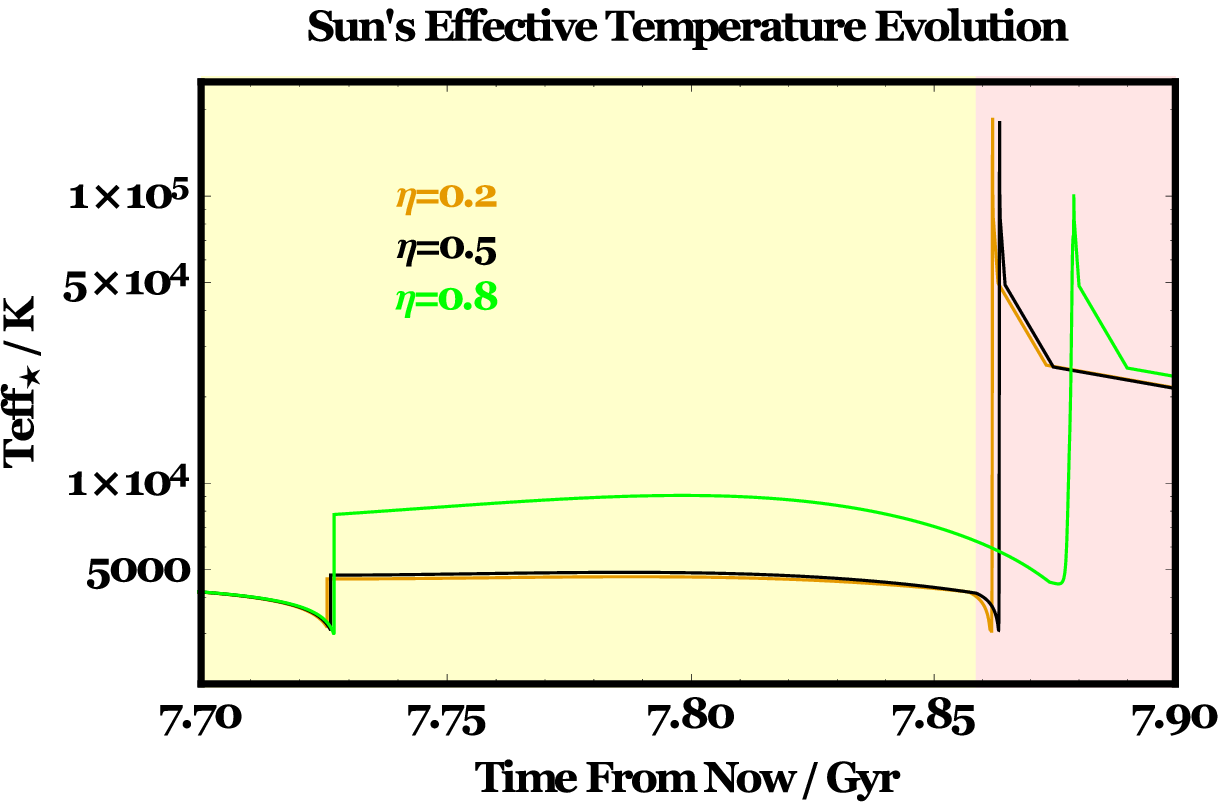}
\includegraphics[width=8cm]{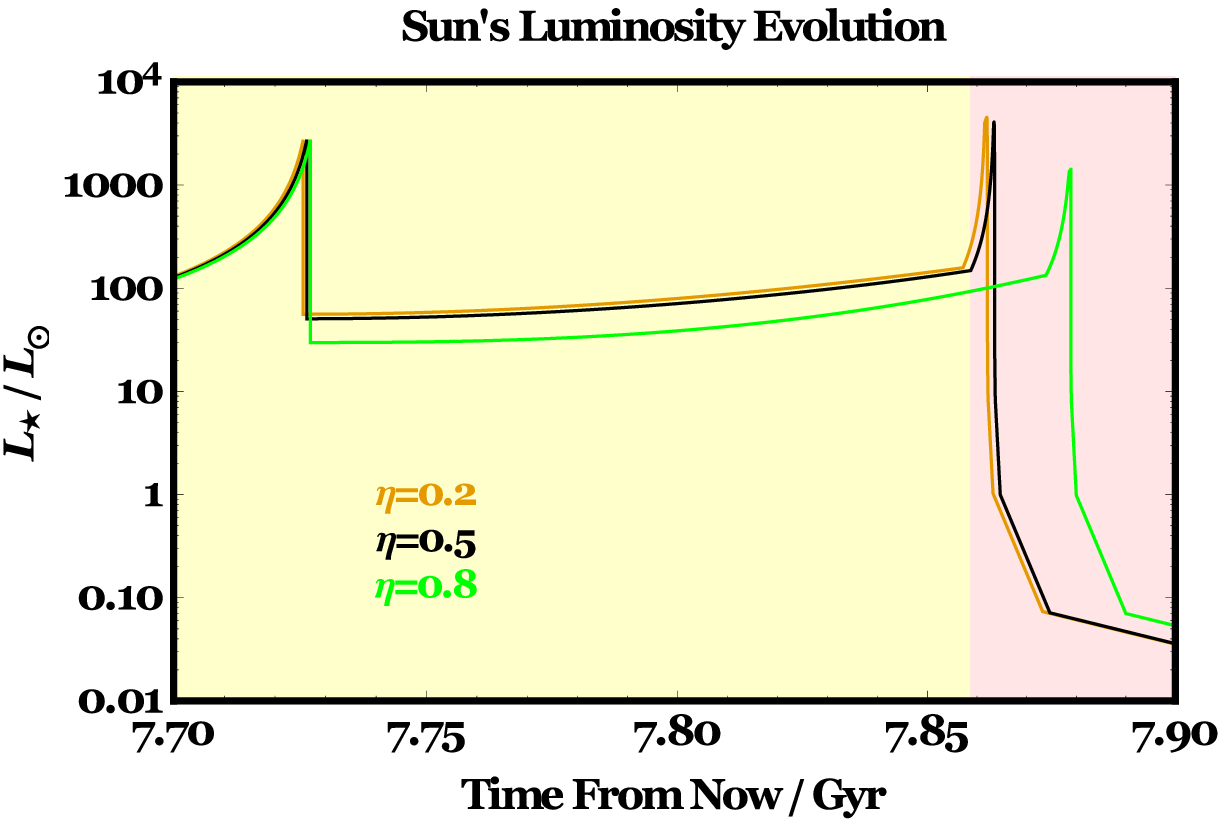}
}
\caption{
The future evolution of the Sun's mass ($M_{\star}$), radius ($R_{\star}$), effective temperature (Teff$_{\star}$) and luminosity ($L_{\star}$) during the transition between the red giant branch phase (yellow background) and the asymptotic giant branch phase (pink background). These evolutionary tracks were computed with the {\tt SSE} code (Hurley et al. 2000).  The $\eta$ parameter (Equation \ref{reimers}) determines the extent of the mass loss on the red giant branch phase, which has implications for evolution on the asymptotic giant branch phase. The phase transition (7.8587 Gyr) is shown for $\eta = 0.5$; other phase transitions are 7.8571 Gyr for $\eta = 0.2$ and 7.8738 Gyr for $\eta = 0.8$. The Sun eventually becomes a white dwarf with a mass of $0.511-0.534M_{\odot}$.
}
\label{sunevo}
\end{figure*}

The Sun will leave the main sequence in about 6.5 Gyr, and undergo drastic changes (Fig. \ref{sunevo}). Its radius will increase by a factor of about 230, it will lose almost half of its current mass, and its luminosity will reach a peak value which is about 4000 times its current value (see e.g \citealt*{schcon2008}, \citealt*{verwya2012} and fig. 3 of \citealt*{veras2016}). The Sun will become so large that its radius will extend just beyond where the Earth currently sits. These major changes will occur in two phases. The red giant branch phase will last about 800 Myr. In this timespan, the Sun will gradually lose about a quarter of its mass. The second phase, when the Sun becomes an asymptotic giant branch star, is quicker: lasting just 5 Myr. Another quarter of the Sun's mass will be lost during this period. During both phases, the radius of the Sun will extend out to nearly the Earth's distance.  

The consequences for the inner Solar system will be profound. The terrestrial planets, which are likely to remain in stable orbits until the end of the main sequence at the approximately $99$\% level \citep{lasgas2009,batetal2015for3,zeebe2015}, will be in danger. Mercury and Venus will be engulfed, and the Earth will be on the edge of survivability \citep{rybden2001,schcon2008}. Mars will be roasted, but should survive, because it will escape being ensnared by the tidal reach of the Sun \citep{villiv2009,kunetal2011,musvil2012,adablo2013,norspi2013,viletal2014,staetal2016}. Asteroid belt constituents between 100~m and 10~km in radius will be spun up to breakup speed \citep{veretal2014afor3}, creating a sea of debris, some of which may be water-rich \citep{jurxu2010,jurxu2012,faretal2013,radetal2015,malper2016}.

The consequences for the giant planets, however, will be more benign. Jupiter, Saturn, Uranus and Neptune will increase their semimajor axes by a factor of about two each, and not undergo scattering nor instability \citep{dunlis1998}, even though the chemistry of at least Jupiter's atmosphere will be fundamentally altered \citep{villiv2007,spimad2012}. The giant planet eccentricities will remain effectively fixed because they reside within the adiabatic limit, beyond which stellar mass loss changes both eccentricity and semimajor axis \citep{veretal2011}. The present-day ``adiabatic''\footnote{In this context, adiabaticity has nothing to do with heat, but rather refers to the conservation of eccentricity.} limit for the Solar system specifically lies between about $10^3$ and $10^4$ au \citep{verwya2012}. Further, even though mass loss changes stability limits \citep{debsig2002,musetal2013for3,portegieszwart2013,veretal2013a,voyetal2013,musetal2014for3,vergae2015,veretal2016a}, this change will not be large enough to affect the giant planets. I perform some simulations here also to back up this statement.

Moons, the Kuiper Belt, scattered disc and Oort Cloud will also be affected. Moons of planets will become more entrenched in the Hill spheres of the host planets, and would stay there in the absence of a planetary scattering event \citep{payetal2016a,payetal2016b}. Although known Kuiper Belt and scattered disc objects are within the adiabatic limit, many are likely to become unstable as the stability limits between Neptune and Kuiper Belt objects change as the Sun's mass decreases (e.g. \citealt*{bonetal2011for3}) and its luminosity increases \citep{veretal2015afor3}. The Oort Cloud will be both excited and decimated \citep{verwya2012,veretal2014b}, which will alter the influx of comets into the inner Solar system \citep{alcetal1986for3,paralc1998,veretal2014cfor3,stoetal2015for3}, and these comets can be subsequently perturbed by radiation from the Solar white dwarf \citep{veretal2015bfor3}.

\subsection{The effect of additional planets}

The presence of additional planets (one or more) can trigger instability purely through Lagrange instability as the Sun becomes older. This phenomenon in main sequence extrasolar systems has been well-explored \citep[e.g.][]{chaetal1996for3,bargre2006,bargre2007,chaetal2008,decetal2013for3,vermus2013,davetal2014,marzari2014,petrovich2015,puwu2015} and will have greater predictive power as asteroseismology continues to pinpoint the ages of old stars \citep{cametal2015for3, siletal2015,veretal2015c,noretal2016}. Mass loss from giant branch evolution will change the limits further, as outlined above. Additionally, a planet residing in the non-adiabatic regime will have its eccentricity changed during this mass loss, altering the limits even further. When combined with Galactic tides and stellar flybys, the result may be a significantly more dynamically active evolved system than previously envisaged.

Hence, although Jupiter, Saturn, Uranus and Neptune by themselves will remain stable throughout the Sun's giant branch and white dwarf phases, the presence of additional planets might trigger instability after the Sun has left the main sequence, but not before. This scenario might be common throughout the Milky Way if extrasolar planetary system architectures from about 5 to 30 au resemble the Solar system's, and contain a sufficiently distant and massive extra planet. In this paper, I quantify this scenario, and place rough bounds on the mass and orbit of one additional planet which can trigger future instability.

\subsection{The polluted planet-host star}

The importance of considering extrasolar systems in general arises from the outcome of the instability during the white dwarf phase. We know that planetary systems around white dwarfs are just as common as those orbiting main sequence stars, primarily through the detection of metal debris in the atmospheres of the degenerate stars. Such ``polluted'' white dwarfs now number in the thousands (\citealt*{dufetal2007,kleetal2013,kepetal2015,kepetal2016} and Hollands et al. in preparation). High-sensitivity observational surveys reveal that between one-quarter and one-half of the known Milky Way white dwarfs are estimated to host planetary debris \citep{zucetal2003,zucetal2010,koeetal2014for3}.

The spectacular discovery of at least one minor planet disintegrating in real time within the white dwarf disruption radius \citep{vanetal2015} has prompted a spate of observational follow-up studies \citep{aloetal2016,croetal2016,ganetal2016,rapetal2016,xuetal2016,zhoetal2016} as well as some theoretical attempts to explain the complex behaviour \citep{guretal2016for3,veretal2016b}. Such minor planets are assumed to supply dusty and gaseous debris discs which orbit the white dwarf \citep{zucbec1987,gaeetal2006,faretal2009for3,beretal2014,wiletal2014,xujur2014,baretal2016,farihi2016,manetal2016}. Such discs eventually accrete onto the atmosphere of the star itself \citep{bocraf2011,rafikov2011a,rafikov2011b,metetal2012for3,rafgar2012,wyaetal2014}.

The specific primary mechanism which transports the rocky bodies to the white dwarf is still debatable (see section 7 of \citealt*{veras2016}) but relies on interactions between minor planets and larger bodies (Bonsor et al. 2011; Bonsor \& Wyatt 2012; Debes, Walsh \& Stark 2012; Frewen \& Hansen 2014).  
Instabilities which arise in white dwarf planetary systems
(Veras et al. 2013a; Voyatzis et al. 2013; Mustill et al. 2014; Veras \& G\"{a}nsicke 2015; Bonsor \& Veras 2015; Payne et al. 2016a,b; Hamers \& Portegies Zwart 2016; Petrovich \& Mu{\~n}oz 2016)
promote an architecture conducive to pollution by placing planets and liberated moons in orbits that sweep through regions of space which may access reservoirs of debris. Examples of such reservoirs are those that arise from planet-planet collisions (Shannon et al. in preparation) or YORP-induced asteroid breakup (Veras et al. 2014a).

Therefore, the ubiquity of white dwarf pollution throughout the Milky Way (Zuckerman et al. 2003, 2010; Koester et al. 2014) suggests that post-main-sequence instability is common, an idea that would be aided by the presence of extrasolar analogues of Planet Nine: i.e. potentially massive planets on wide orbits that remain stable during the main sequence phase but begin to scatter gravitationally during or after giant branch mass loss. If the presence of planets at hundreds or thousands of au is the norm, regardless of their dynamical origin, then that scenario is broadly consistent with the observed end states of planetary systems.

\subsection{Outline for this paper}

In this paper I characterize the mass and orbital parameters of an additional distant planet that would create gravitational instability amongst giant planet analogues orbiting a Sun-like star after the latter turns off of the main sequence. This result would also apply to the Solar system if no Planet Nine is found now: at some point later in the Sun's main sequence lifetime the Sun may capture an additional planet \citep{sumetal2011,perkou2012,varetal2012for3}.

In Section 2, I provide more detail about the evolution of Sun-like stars, before, in Section 3, isolating the varied effects that come into play during the giant branch and white dwarf phases. My computational method is detailed in Section 4. I explain the initial conditions and report on my simulation results in Section 5 before summarizing the conclusions of this work in Section 6.

\section{The Sun's future evolution}

Amongst the Sun's many physical properties (such as radius and luminosity), its mass is the most important one for my study. On the main sequence, the Sun's mass loss rate ($\approx 2.4 \times 10^{-14} M_{\odot}$ yr$^{-1}$) is negligible \citep{vial2013}. However, the Sun's future evolution beyond the main sequence is unknown. A potentially good guess is to consider the typical evolution of other $1M_{\odot}$ stars with metallicities of $Z = 0.02$. Although these two parameters are well-established for the Sun, an important unknown is the future rate of mass loss during the red giant branch phase.

A traditional formulation of this rate is the Reimers prescription \citep{reimers1975,reimers1977}, which has now been improved \citep{schcun2005}, and may be parametrized as

\[
\frac{dM_{\star}}{dt} =\eta \left(4 \times 10^{-13} M_{\odot} \ {\rm yr}^{-1} \right) 
\left( \frac{L_{\star}}{L_{\odot}} \right)
\left( \frac{R_{\star}}{R_{\odot}} \right)
\left( \frac{M_{\star}}{M_{\odot}} \right)^{-1}
\]

\begin{equation}
\ \ \ \ \ \ \ \ \ \times
\left( \frac{T_{\star}}{4000 \ {\rm K}} \right)^{7/2}
\left[1 + 2.3 \times 10^{-4} \left( \frac{g_{\star}}{g_{\odot}} \right)^{-1} \right]
,
\label{reimers}
\end{equation}

\noindent{}where $M_{\star}$, $L_{\star}$, $R_{\star}$, $T_{\star}$ and $g_{\star}$ 
are the mass, luminosity, radius, temperature and surface gravity of the star.  
The value of $\eta = 0.2$ 
reproduces the coefficient given
in \cite{schcun2005}, and $\eta = 0.5$ reproduces the coefficient given
in the traditional formulation, which does not contain the terms dependent
on $T_{\star}$ nor $g_{\star}$.

Equation (\ref{reimers}) applies to the giant branch phase only.
Mass loss evolution on the asymptotic giant branch phase is qualitatively different,
and is typically characterised by the prescription of \cite{vaswoo1993}:

\begin{equation}
\log{\left(\frac{dM_{\star}}{dt}\right)} = -11.4 + 0.0125 
\left[P - 100 {\rm max}\left(M_{\star} - 2.5, 0.0 \right)  \right] 
\end{equation}

\noindent{where} $dM_{\star}/dt$ is computed in $M_{\odot}$ yr$^{-1}$ and such that

\begin{equation}
\log{P} \equiv {\rm min}\left(3.3, \ -2.07 - 0.9 \log{M_{\star}} + 1.94 \log{R_{\star}} \right)
,
\end{equation}

\noindent{where} the value $P$ is computed in years. The 
continued common
use of this prescription even over two decades after it was published helps indicate
its robustness in the face of new observations. This formulation also importantly
includes the ``superwind'' (peak mass loss) which occurs at the ``tip'' (end-point) 
of the asymptotic giant branch \citep{lagzij2008}.

After the asymptotic giant branch phase, the Sun will become a white dwarf.
The Solar white dwarf will not lose mass nor change its radius, but will gradually dim. 
Figure \ref{sunevo} illustrates the Sun's evolution for ($\eta = 0.2, 0.5, 0.8$)
from the {\tt SSE} code, which uses the traditional Reimers formulation and 
will be described in more detail in Section 4.
I adopted these values in order to encompass a realistic range, as outlined above, 
and further to conform to the range adopted in \cite{verwya2012}.
Fig. \ref{sunevo} displays the Sun's
mass evolution during the transition between the red giant branch and 
asymptotic giant branch for all three values (also included for added perspective are the Sun's
radius, luminosity and effective temperature changes). The resulting Solar white dwarf mass for 
each of these values is $0.534 M_{\odot}$ ($\eta=0.2$), 
$0.519 M_{\odot}$ ($\eta=0.5$) and $0.511 M_{\odot}$ ($\eta=0.8$). 

\begin{figure*}
\centerline{
\includegraphics[width=8cm]{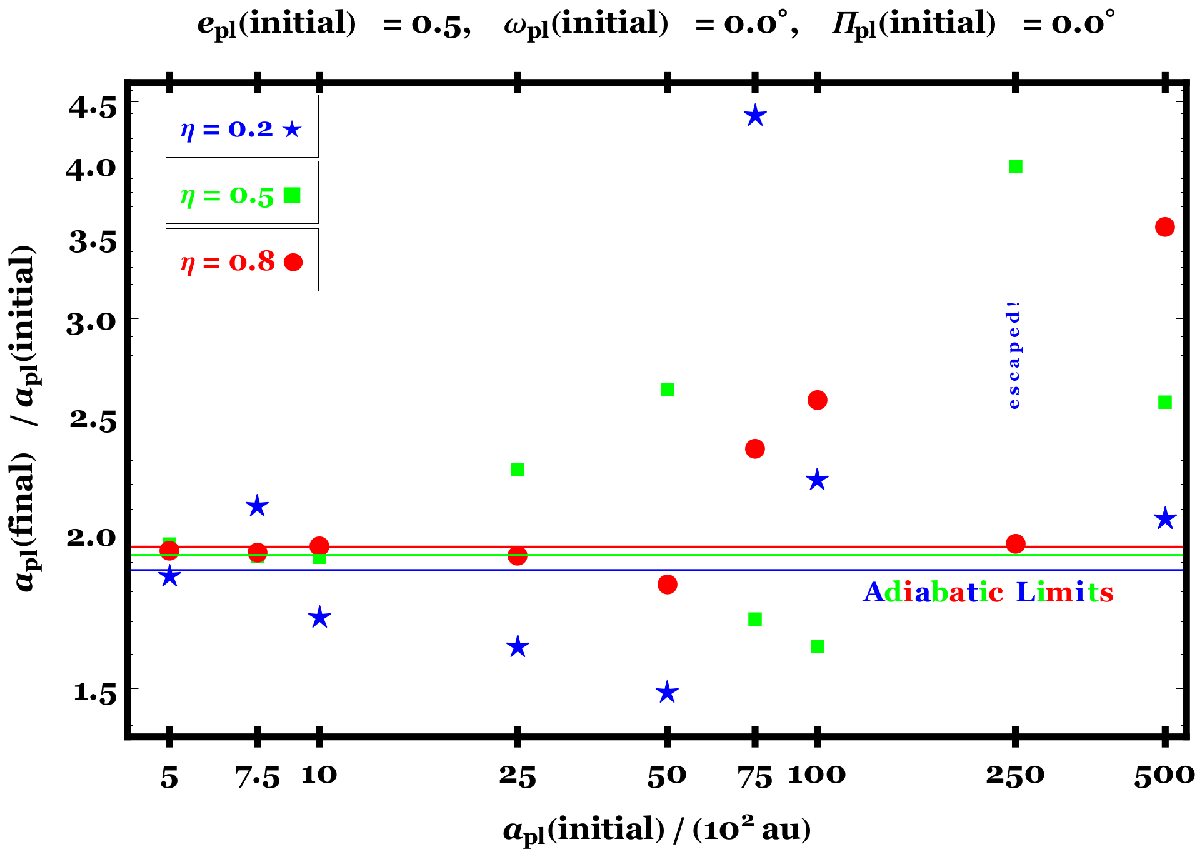}
\includegraphics[width=8cm]{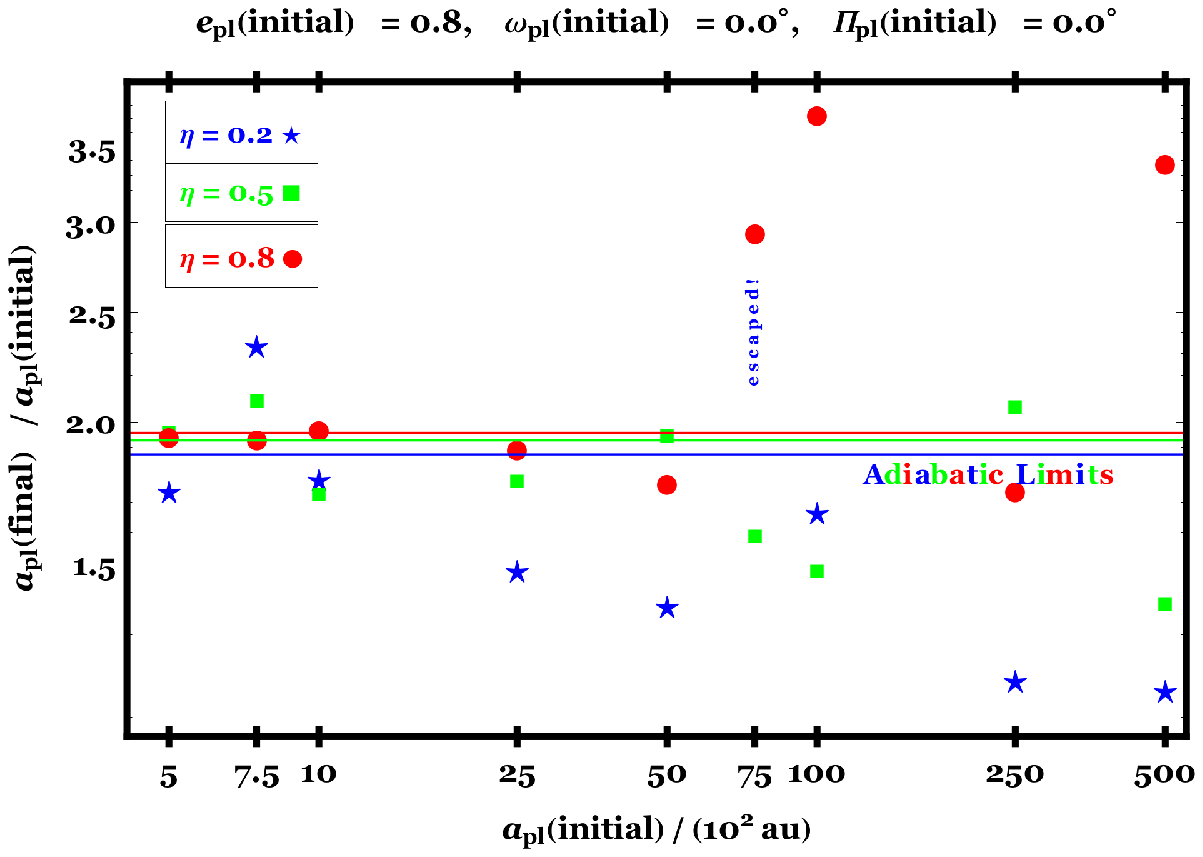}
}
\
\
\centerline{
\includegraphics[width=8cm]{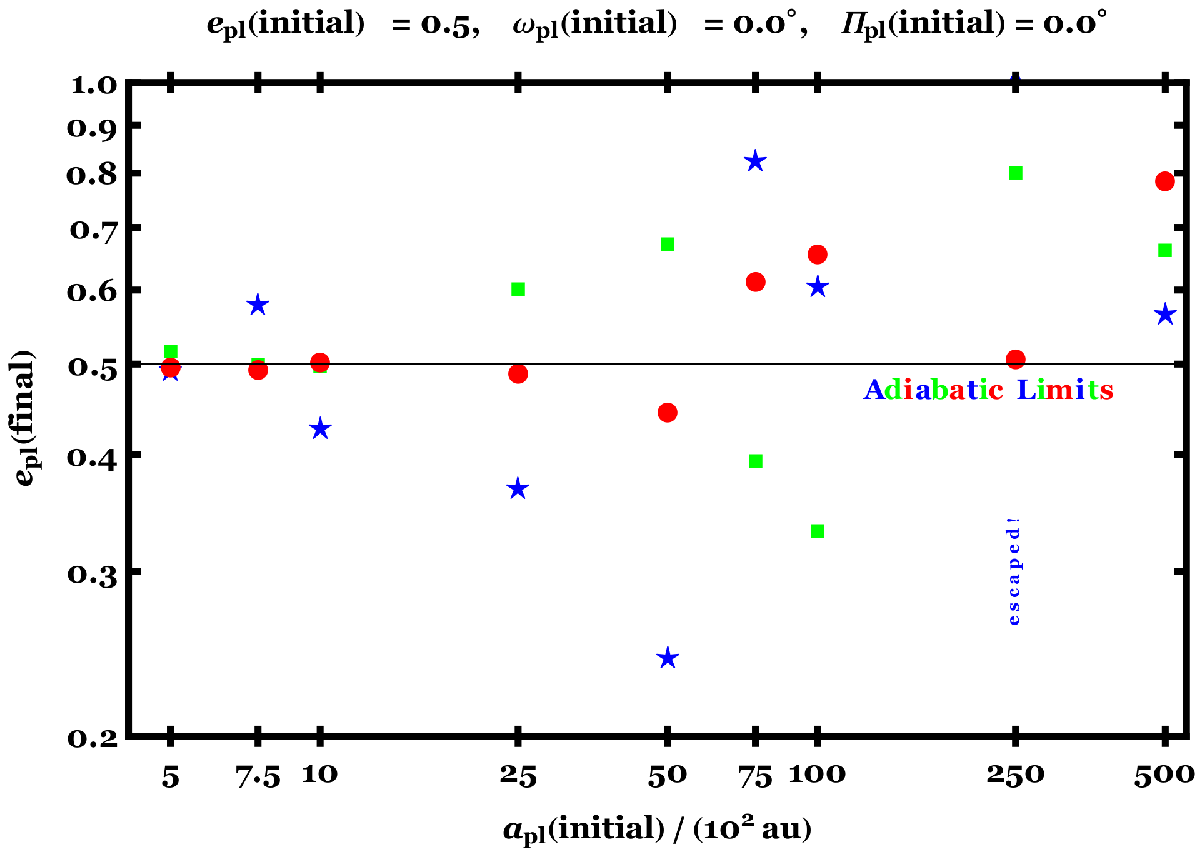}
\includegraphics[width=8cm]{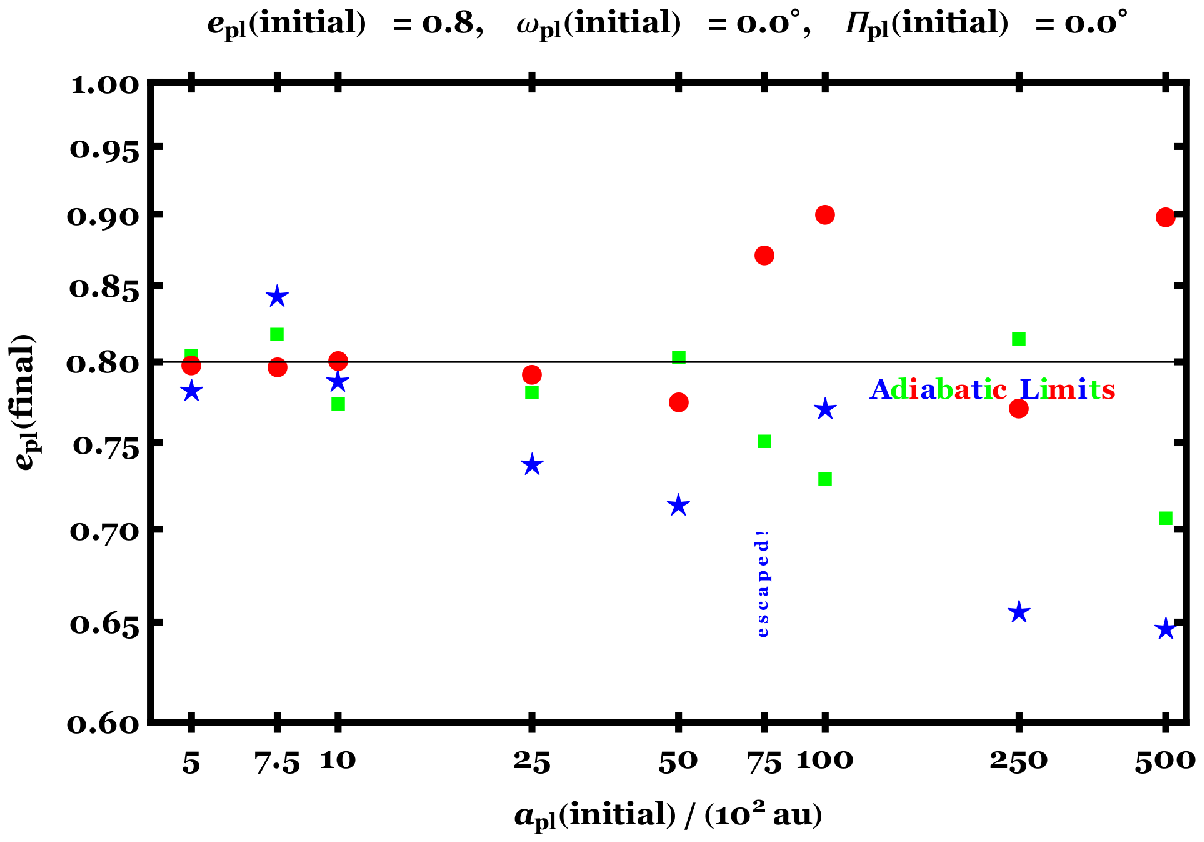}
}
\caption{
The final semimajor axis and eccentricity of an isolated planet orbiting
the Solar white dwarf, assuming no other Solar system planets exist and the only
active force was and is gravity from the Sun.
The left panels show the case for a main sequence, or, ``initial'' $e_{\rm pl} = 0.5$ 
value and the right panels for an initial $e_{\rm pl} = 0.8$. The adiabatic limits, given 
by the straight horizontal lines, are reliable predictors typically only for 
$a_{\rm pl} \ll 500$ au (specifically see Eq. \ref{mlindex}).
These plots illustrate the non-monotonicity of the planet's evolution
as a function of initial $a_{\rm pl}$.
}
\label{adiasemi}
\end{figure*}

\begin{figure*}
\centerline{
\includegraphics[width=8cm]{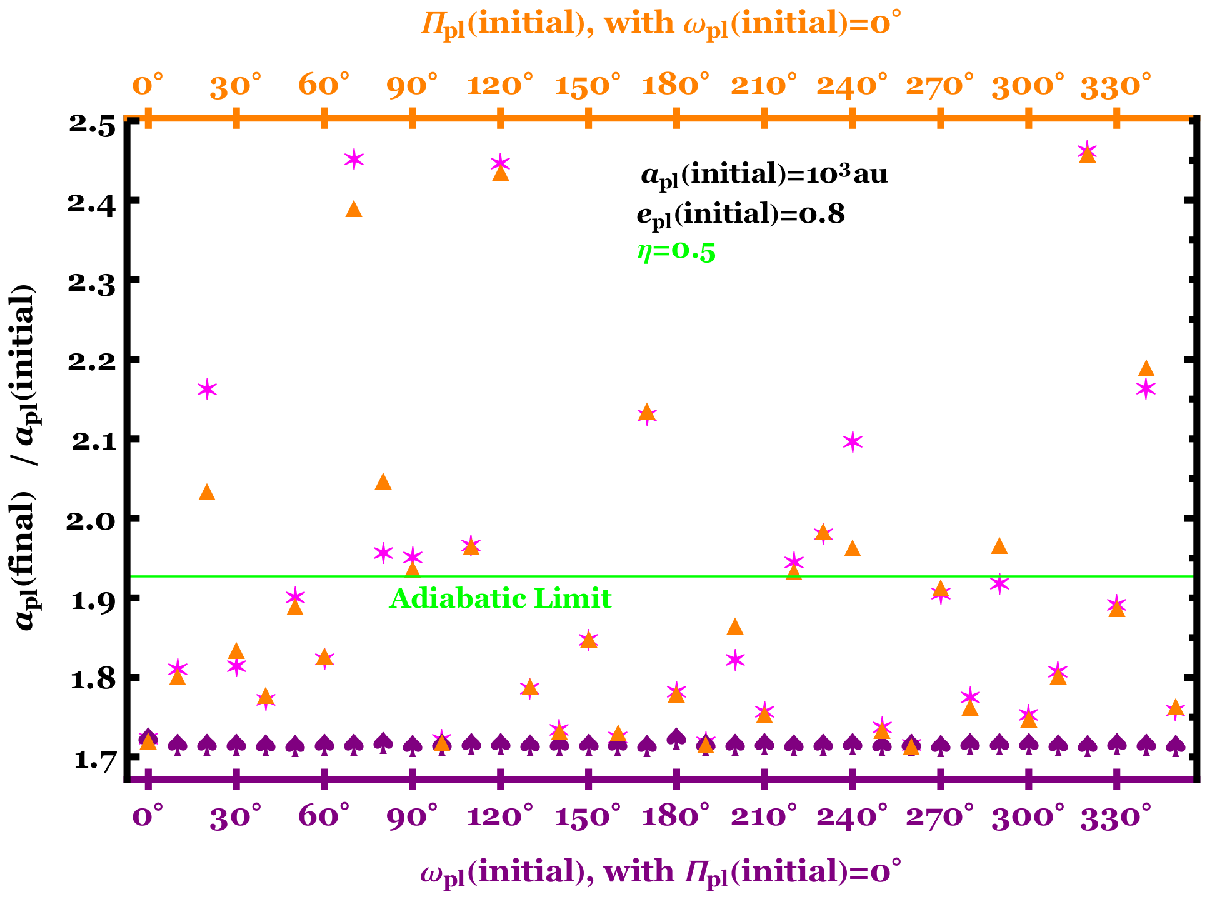}
\includegraphics[width=8cm]{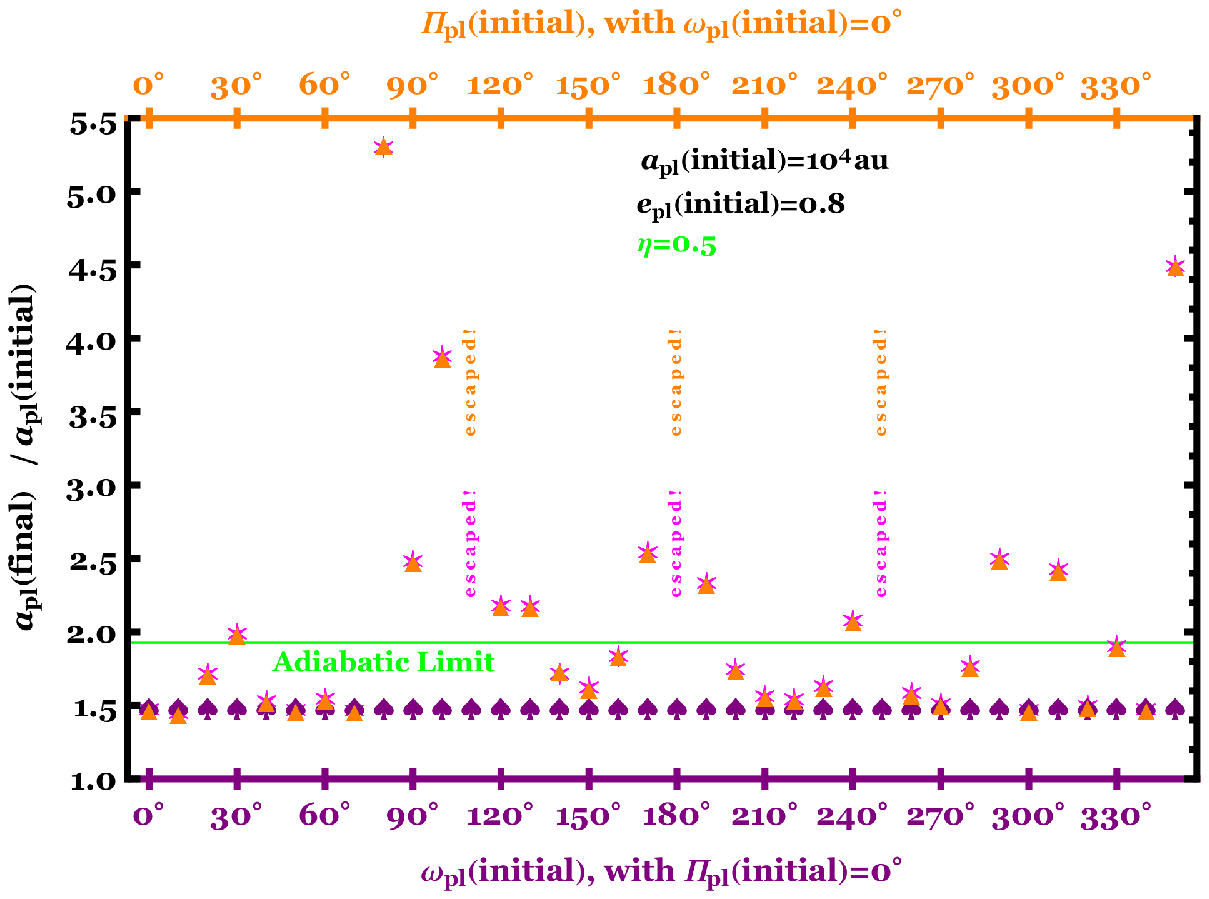}
}
\
\
\centerline{
\includegraphics[width=8cm]{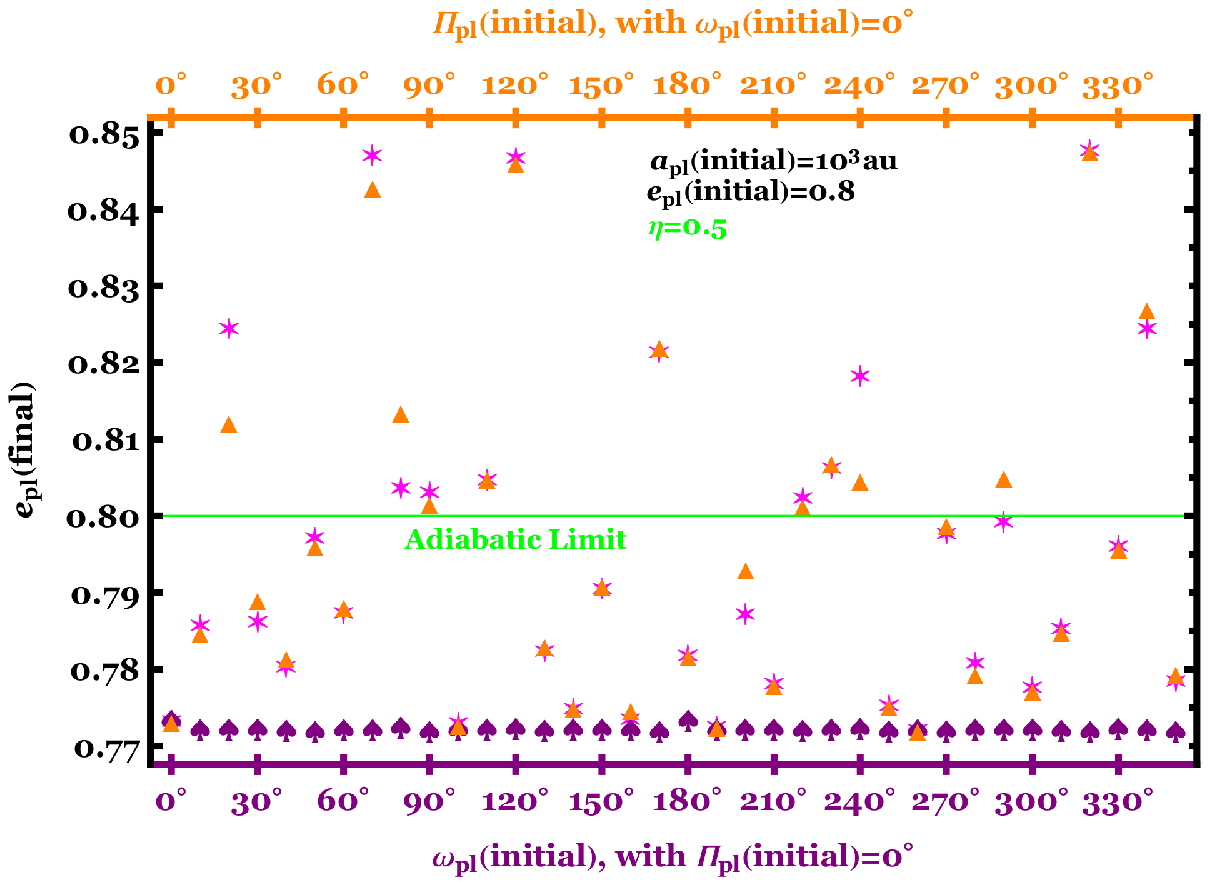}
\includegraphics[width=8cm]{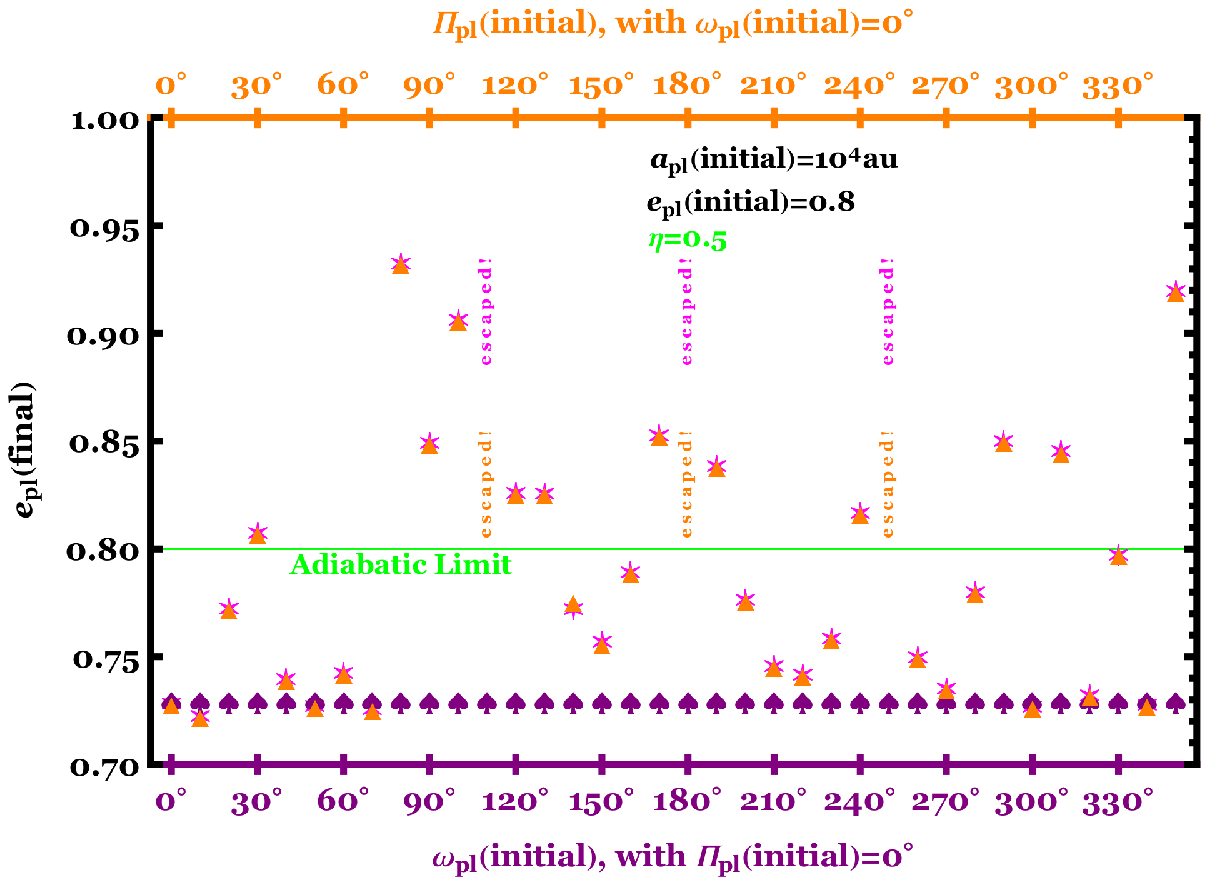}
}
\caption{
An isolated planet's evolution dependence on physical location
and argument of pericentre. 
This figure is similar to Fig. \ref{adiasemi},
except here I did not vary the main sequence, or initial, values of $a_{\rm pl}$ and
$e_{\rm pl}$ ({\it left panels}: $a_{\rm pl} = 10^3$ au and $e_{\rm pl} = 0.8$ and
{\it right panels}: $a_{\rm pl} = 10^4$ au and $e_{\rm pl} = 0.8$) 
while instead varying
$\omega_{\rm pl}$ (argument of pericentre) and $\Pi_{\rm pl}$ (mean anomaly). 
The purple spades and orange triangles
respectively demonstrate the effect of varying initial $\omega_{\rm pl}$
and $\Pi_{\rm pl}$. The difference is dramatic, and helps exhibit the
sensitive dependence of mass loss on planet position along the orbit.
This sensitivity is further
emphasized by the pink stars, which result from
the initial values of $\omega_{\rm pl}$ and $\Pi_{\rm pl}$ being equal to
each other and to the
$x$-axes values. Because the pink stars and orange triangles become
more coincident as $a_{\rm pl}$ is increased (from the left panels to the right panels),
non-adiabaticity directly correlates with sensitivity on $\Pi_{\rm pl}$,
a known theoretical result (Veras et al. 2011). The larger value of $a_{\rm pl}$
also allows for some systems to become unbound.
}
\label{adiaangles}
\end{figure*}

\section{Isolating different effects}

The motion of any planet orbiting a star at a distance of at least hundreds or thousands of au 
could be affected by a variety of effects. These include post-main-sequence mass loss, 
Galactic tides, and flybys from passing stars
(see fig. 2 of \citealt*{veras2016}). Combining these effects with the gravitational interactions 
from a set of giant planets like Jupiter, Saturn, Uranus and Neptune is non-trivial,
especially during close encounters.
I am interested in the consequences of these effects on the orbits of giant planets,
and in the below subsections, isolate these physical processes.

Consider first a single planet with an arbitrary sub-stellar mass
in isolation on an arbitrary orbit whose pericentre lies beyond $10^2$ au from
the Sun-like star.

\subsection{Stellar flybys}

Stars pass by the Sun on a regular basis, and occasionally enter the Sun's gravitational
sphere of influence. Sometimes these intrusions are deep. In fact, 
a Sun-like star can expect to encounter another star at a distance of a few hundred
au over its main sequence lifetime \citep{zaktre2004,vermoe2012}. 
The consequences for a distant planet may be any outcome
(orbit perturbation, collision, ejection, engulfment)
depending on the velocity, direction and mass of the flyby star. On Gyr-timescales,
objects with semimajor axes of the order $10^4$ au have
a significant chance of being stripped away by passing stars \citep{liada2016}.

These flybys will occur during all phases of stellar evolution. However, predicting where and 
when these flybys will occur is generally not possible
because the stellar kinematic memory in the Galaxy is lost well-before 1 Gyr in the future
\citep{naketal2010for3}. In fact, the significant distance errors on the closest known Solar flyby, 
which occurred about $70_{-15}^{+10} \times 10^3$ yr ago, are $52_{-14}^{+23} \times 10^3$ au 
\citep{mametal2015}. Consequently, accurately predicting the incidence of flybys on the 
giant branch or white dwarf phases of Sun-like stars is not possible, and hence I do
not consider flybys any further.

\subsection{Galactic tides}

The effect of Galactic tides is more predictable. In the Solar 
neighbourhood\footnote{I did not consider the possibility that the parent star 
has migrated or will radially migrate   
within the Galaxy \citep{selbin2002,rosetal2008}.}, the
vertical component of these tides is an order of magnitude stronger than the planar
components \citep{He86,Ma89,Ma92,Ma95,Br96for3,Br01,Br05}, greatly simplifying the 
equations of motion \citep[see the contrast within table 1 of][]{veretal2014b}.
For most orbits except those near the escape boundary, Galactic tides change
a planet's eccentricity $e$, inclination $i$, argument of periastron $\omega$ and 
longitude of ascending node $\Omega$
-- but not its semimajor axis $a$ \citep{Br01,Fo04,Fo06,vereva2013a}. However,
near the escape boundary (within a factor of 3-10 in orbital period), all
of the orbital elements vary \citep[fig. 3 of][]{veretal2014b}.

The strength of these tidal changes depends primarily on the inclination of the planetary orbit
with respect to the Galactic plane, with the maximum effect achieved for polar orbits.
In this extreme case, the planet's eccentricity may reach unity and consequently escape 
the system. In the other extreme, coplanar, case
the minimum eccentricity increase and decrease factors generated by Galactic tides
in the Solar neighbourhood are 
roughly $1 \pm 10^{-8}(a/{\rm au})^{3/2}$ \citep[eq. 44 of][]{vereva2013a}. This value
corresponds to an eccentricity change of about $10^{-5}$ for $a = 10^2$ au and 
$10^{-2}$ for $a = 10^5$ au.

The Solar system, however, does not fit either extreme. The ecliptic is inclined to the 
Galactic plane by about $60^{\circ}$, and the orbits of the outer planets therefore 
have a relative inclination of about $57.5^{\circ}-62.5^{\circ}$ with respect to the
Galactic plane. This orientation dictates that an object like Sedna (with a semimajor
axis of about 500 au and eccentricity of 0.86) 
will change its eccentricity by about $10^{-3} - 10^{-2}$ from now until the end of the main
sequence \citep[fig. 2 of][]{vereva2013b}. At $10^3$ au, a substellar companion may significantly
change its eccentricity by $10^{-1}$ after 10 Gyr \citep[fig. 3 of][]{vereva2013a}.

Overall then, the inclination of a distant planet with respect to the Galactic plane
may importantly be affected by Galactic tides, with a potential subsequent effect on the
stability of the remaining planets in the system.

\subsection{Stellar mass loss}

Stellar mass loss will likely represent the key dynamical driver of the future Solar
system, as well as for most exoplanetary systems.  The motion of a distant
planet will be determined by the solution of the following set
of differential equations \citep{omarov1962,hadjidemetriou1963,veretal2011}

\begin{eqnarray}
\frac{da_{\rm pl}}{dt} &=& 
-\frac{a_{\rm pl} \left(1 + e_{\rm pl}^2 + 2 e_{\rm pl} \cos{f_{\rm pl}}\right)}{1 - e_{\rm pl}^2}
\frac{\dot{M}_{\star} + \dot{M}_{\rm pl}}{M_{\star} + M_{\rm pl}}
\\
\label{eqml3af}
&=&
-a_{\rm pl}\left(\frac{1+e_{\rm pl}\cos{E_{\rm pl}}}{1 - e_{\rm pl}\cos{E_{\rm pl}}} \right)
\frac{\dot{M}_{\star} + \dot{M}_{\rm pl}}{M_{\star} + M_{\rm pl}} 
,
\label{eqml3a}
\\
\frac{de_{\rm pl}}{dt} &=& -\left(e_{\rm pl} + \cos{f_{\rm pl}} \right)
\frac{\dot{M}_{\star} + \dot{M}_{\rm pl}}{M_{\star} + M_{\rm pl}} 
\ \ \ \ \ \ \ \ \ \ \ \ \ \ \ \ \ \ 
\\
&=&
-\left[ \frac{\left(1 - e_{\rm pl}^2\right) \cos{E_{\rm pl}}}{1 - e_{\rm pl} \cos{E_{\rm pl}}} \right]
\frac{\dot{M}_{\star} + \dot{M}_{\rm pl}}{M_{\star} + M_{\rm pl}}
,
\label{eqml3e}
\\
\frac{di_{\rm pl}}{dt} &=& 0
,
\label{eqml3i}
\\
\frac{d\Omega_{\rm pl}}{dt} &=& 0
,
\label{eqml3bigO}
\\
\frac{d\omega_{\rm pl}}{dt} &=& -\left(\frac{\sin{f_{\rm pl}}}{e_{\rm pl}}\right) \frac{\dot{M}_{\star} + \dot{M}_{\rm pl}}{M_{\star}+M_{\rm pl}} 
\\
&=&
-\left[ \frac{\sqrt{1-e_{\rm pl}^2} \sin{E_{\rm pl}}}{e_{\rm pl} \left(1 - e_{\rm pl} \cos{E_{\rm pl}} \right) } \right]
\frac{\dot{M}_{\star} + \dot{M}_{\rm pl}}{M_{\star} + M_{\rm pl}}
,
\label{eqml3smallo}
\end{eqnarray}

\noindent{}where $f_{\rm pl}$ and $E_{\rm pl}$ refer to the true anomaly and eccentric anomaly
of the planet.
I have denoted the star's (changing) mass as $M_{\star}$ ($M_{\odot}$
is just the current value of the mass of a Sun-like star) and the
planet's mass as $M_{\rm pl}$. 
The evolution of the anomalies is

\begin{eqnarray}
\frac{df_{\rm pl}}{dt} &=& -\frac{d\omega_{\rm pl}}{dt} + \frac{n_{\rm pl} \left(1 + e_{\rm pl} \cos{f_{\rm pl}}\right)^2}{\left(1 - e_{\rm pl}^2\right)^{\frac{3}{2}}}
,
\label{eqml3f}
\\
\frac{dE_{\rm pl}}{dt} &=& -\frac{1}{\sqrt{1-e_{\rm pl}^2}} \frac{d\omega_{\rm pl}}{dt} + \frac{n_{\rm pl}}{1 - e_{\rm pl} \cos{E_{\rm pl}}}
.
\label{eqml3E}
\end{eqnarray}

\noindent{}where the mean motion
$n_{\rm pl} = G^{1/2}\left(M_{\star} + M_{\rm pl}\right)^{1/2}a_{\rm pl}^{-3/2}$, and generally, $\dot{M}_{\star} < 0$. Because the planet accretes some of the stellar ejecta (\citealt*{hadjidemetriou1963} and section 4.2.2. of \citealt*{veras2016})
$\dot{M}_{\rm pl} > 0$. However, for Solar system analogues, $\dot{M}_{\rm pl}/\dot{M}_{\star} \ll 1$. 

Equations (\ref{eqml3af})-(\ref{eqml3E}) are not known to be solvable analytically,
and assume that the Sun-like star loses mass isotropically.
\cite{veretal2013bfor3} showed that this approximation is excellent because 
latitudinal mass-loss variations anisotropically affect the planet's motion only if
the mass loss is asymmetric about the stellar equator. Unrealistic longitudinal variations
(of over 0.1\%) must be maintained for long periods (over 1 Myr) in order to produce
semimajor axis variations of order 0.1 au and eccentricity variations of order 0.01. See 
their fig. 2 for rough magnitudes, although their choice of reference frame alters the interpretation of the orbital elements \citep{doskal2016a,doskal2016b}. 

The smaller the value of $a_{\rm pl}$, the better that the semimajor axis evolution can be approximated in the adiabatic limit as

\begin{equation}
\left( \frac{da_{\rm pl}}{dt} \right)_{\rm adiabatic} = -a_{\rm pl}\left(\frac{\dot{M}_{\star} + \dot{M}_{\rm pl}}{M_{\star} + M_{\rm pl}}\right)
.
\end{equation}

\noindent{}The goodness of this approximation can be quantified by the time variable mass loss index $\Psi$, introduced by equation 15 of \cite{veretal2011} as

\begin{eqnarray}
\Psi &\equiv& \frac{\dot{M}_{\star} + \dot{M}_{\rm pl}}{n_{\rm pl} \left(M_{\star} + M_{\rm pl}\right)}
\nonumber
\\
&\approx& 0.005 
\left( \frac{\dot{M}_{\star} + \dot{M}_{\rm pl}}{10^{-6} M_{\odot}/{\rm yr}}\right)
\left( \frac{a_{\rm pl}}{10^3 \ {\rm au}}\right)^{\frac{3}{2}}
\left( \frac{M_{\star} + M_{\rm pl}}{1 M_{\odot}}\right)^{-\frac{3}{2}}
.
\label{mlindex}
\end{eqnarray}

\noindent{}The ``adiabatic regime'' is a useful characterisation for the regime which 
typically occurs when $\Psi \ll 0.1-1$. Consequently, a planet located at $10^3$ 
au should be approaching the edge of the
adiabatic limit. In order to quantify this sentiment, I have run simulations
of a planet at different semimajor axes and two different eccentricities,
for each of the three $\eta = 0.2, 0.5, 0.8$ stellar models. Fig. \ref{adiasemi}
illustrates the results.

The figure demonstrates how non-adiabaticity increases with initial (or, main sequence values
of) $a_{\rm pl}$. For
$\eta = 0.8$, which has the slowest rate of mass loss of the three tracks that I sampled
despite releasing the most mass overall (see Fig. \ref{sunevo}), the adiabat is followed 
most closely (within a few per cent) for distances including and under $10^3$ au.
In all cases, the departure from adiabaticity is non-monotonic with increasing $a_{\rm pl}$,
and one system in the left panels and one in the right become unbound. Such behaviour
is explained further in \cite{veretal2011} and \cite{verwya2012}.

The systems sampled in Fig. \ref{adiasemi} fix the planet's initial argument of pericentre
and mean anomaly ($\equiv \Pi_{\rm pl}$). However, the value of these orbital elements is
potentially of crucial importance. Hence, in Fig. \ref{adiaangles}, I fixed the initial
values of $a_{\rm pl}$ and $e_{\rm pl}$ and instead varied orbital angles. The almost straight lines of
purple spades in all of the plots reveal that (i) this choice produces a notable deviation
from the adiabatic limit, and (ii) for the particular initial fixed $\Pi_{\rm pl}$
value of $0^{\circ}$, changing the initial value of $\omega_{\rm pl}$ has little effect on the overall
outcome. Conversely, the scattered orange triangles and pink stars demonstrate a strong
dependence on the planet's location along its orbit during the Solar giant branch phase.
As shown by the right panels, if the initial $a_{\rm pl}$ is large enough, then
values of $\Pi_{\rm pl}$ may be found which would lead to escape.

In both Figs \ref{adiasemi} and \ref{adiaangles} the mass of the planet is not reported.
The reason is that its mass is largely insensitive to the process of mass loss.
In particular, because $M_{\rm pl}$ always appears in summation 
with $M_{\star}$ in equations (\ref{eqml3af})-(\ref{mlindex})
for all $M_{\rm pl} \ll M_{\odot}$, the influence of planet mass on its resulting motion
should enter at the $0.1 \%$ level for Jupiter-mass planets.

\subsection{Galactic tides and stellar mass loss together}

The interplay between Galactic tides and mass loss can be partitioned
because each acts on a different timescale. \cite{veretal2014b} concluded
that in the Solar neighbourhood the tidal timescale is orders of magnitude 
longer than the mass
loss timescale, allowing one to decouple the equations of motion
in each phase. In other words, one can use the tidal equations along
the main sequence and white dwarf phases, and the mass loss equations alone
along the giant branch phases.

The consequences for long-term stability are highlighted in \cite{bonver2015}.
Consider a wide-orbit ($>$~$10^3$ au) planet whose pericentre during
the main sequence is high enough to not perturb an inner planetary system.
After mass loss, the orbit expands enough to cause Galactic tides to
excite the distant planet's eccentricity to the extent that the inner planetary 
system is now affected (see, in particular, their fig. 2). The eccentricity of the distant
planet is likely to further change if during mass loss the planet is in the
non-adiabatic regime
(see Figs \ref{adiasemi} and \ref{adiaangles}).

\subsection{Planet-planet scattering}

By themselves, Jupiter, Saturn, Uranus and Neptune are not assumed
to scatter off one another during the Sun's evolution.
\footnote{Technically, at some point in the distant future (e.g.
many Hubble times) the system will become unstable \citep[e.g.][]{murhol1997}.}
In order to confirm this notion, I have performed three simulations
($\eta = 0.2, 0.5, 0.8$) with the four giant planets only plus Galactic tides.
I ran these simulations throughout the giant branch phases and for 10 Gyr
on the white dwarf phase. In no case did the giant planets commence strong
scattering.

Now consider the inclusion of a distant planet.
Although unimportant for Galactic tides or
stellar mass loss, the mass of the planet will crucially determine 
whether Jupiter, Saturn, Uranus
and Neptune will be perturbed enough to cause large-scale instability
in the Solar system analogue. A distant planet which is comparable in mass to any 
of the known gas giants will likely create such an instability during a close
encounter. What happens if the planet is a super-Earth-mass
planet is less clear. Before exploring these possibilities,
I describe in detail the computational method used in this work
and lay out the initial conditions for my simulations.

\section{Computational method}

\subsection{Numerical codes}

In order to model the future evolution of a Sun-like star concurrently with
the dynamical evolution
of Jupiter, Saturn, Uranus, Neptune analogues and a distant planet,
I utilized a facility which
has combined the stellar evolution
code {\tt SSE} (Hurley et al. 2000) with a heavily modified version of the 
Bulirsch-Stoer integrator in the planetary dynamics code {\tt Mercury} \citep{chambers1999}. 
This combination has been shown to yield converging errors as
the accuracy parameter decreases \citep[e.g. fig. 1 of][]{veretal2013a}, and the resulting
code has been used in several previous studies \citep[e.g.][]{veretal2016a}.
I adopt the version of the code that includes effects from Galactic tides, and 
a realistic Hill ellipsoid, in order to correctly track escape 
\citep{vereva2013a,veretal2014b}. For these computations, I assumed that the
Sun-like star resides at exactly 8~kpc away from the Galactic Centre.

The Hill ellipsoid is stretched towards the centre of the Galaxy,
and compressed along the other two axes. At a distance of 8~kpc,
the semi-axes of this ellipsoid are $(1.49, 1.92, 2.89) \times 10^5$ au.
Any planet which the code found to reside outside of this ellipsoid
at any timestep was considered to have escaped from the planetary system.
None of the $M_{\rm pl}$ values that I chose were large enough to alter these
semi-axes.

The {\tt SSE} code produces a single evolutionary track given 
a set of zero-age-main-sequence values of $M_{\star}$, $Z$ and $\eta$.
As previously mentioned, $\eta$ is a key unknown (Fig. \ref{sunevo}).
Rather than adopting a single value, I chose three ($0.2, 0.5, 0.8$)
for the reasons outlined in Section 2.

\subsection{Simulation duration}

The timespan over which I ran the simulations was limited by 
computational resources because of the number of orbits that 
needed to be accurately modelled. This value is dependent on the
parent star's mass and size of the orbits. Hence, evolution during the
white dwarf phase can proceed at a rate {\it tens of times} faster 
\citep[fig. 8 of][]{veretal2013a} than on the main sequence. 

This computational slowness along the main sequence (together with the difficulty
of self-consistently modelling multiple phases of stellar evolution) partly illustrates
why many post-main-sequence exoplanetary
evolution studies have restricted their simulations to a timespan of under about
1 Gyr
(Debes \& Sigurdsson 2002; Bonsor et al. 2011; Debes et al. 2012; Kratter \& Perets 2012;
Frewen \& Hansen 2014).
Although multi-Gyr simulations were later performed by \cite{veretal2013a} and
Mustill et al. (2014), they had to restrict the main sequence
stellar masses which they sampled to $M_{\star} \ge 3M_{\odot}$ because of the correspondingly
short main sequence lifetimes ($\lesssim 500$ Myr).  Nevertheless, \cite{vergae2015} 
and \cite{veretal2016a} achieved full-lifetime (14 Gyr) simulations by adopting stars 
with main sequence masses between $1.5-2.5M_{\odot}$, and hence main sequence lifetimes of
600 Myr to 3 Gyr.

However, the main sequence lifetime for a $1.0M_{\odot}$ star is much longer, and
approaches 11 Gyr,
meaning that the Sun will remain on the main sequence for another 6.5 Gyr
or so. This timespan
was too prohibitive for my numerical integrations. Consequently, I
began all my simulations just before the red giant branch phase. Nevertheless,
I ran these simulations for a total of 11.6 Gyr, which included roughly 1.5 Gyr 
on the giant branch phases, and 10 Gyr on the white dwarf phase. In order to
maximize resources, the occurrence of instability (defined as ejection or
collision) terminated a simulation. I ran the simulations
which generated Figs \ref{adiasemi} and \ref{adiaangles} for just 1.55 Gyr,
because only the giant branch phases needed to be sampled.

\subsection{Planet parameters}

Regarding the planets, as mentioned previously, Jupiter, Saturn, Uranus and
Neptune are expected to survive until the end of the main sequence in approximately
their current orbits.
Their exact positions 6.5 Gyr from now are not known \citep[e.g.][]{zeebe2015},
even if the errors in current measurements were as small as a Planck length,
because the outer Solar system itself is chaotic with a Lyapunov time of order 10 Myr.
Hence, I simply adopted as
initial conditions the values of $a$, $e$ and $f$ of the planets at the Julian Date 
2451000.5 that are provided within the default version of {\tt Mercury}. 

However, I treated their values of $i$, $\Omega$, and $\omega$ differently.
The code considers the Galactic Plane to exist at $i=0^{\circ}$, and the relative inclination
of each planet to that plane may be important when coupled with the distant planet's 
inclination and the effect of Galactic tides. Because the ecliptic is
approximately inclined at an angle of $60^{\circ}$ to the planets, I
reproduced faithful analogues by tilting the planetary systems so that
the orbital planes are approximately centred about this value. I did so by
applying the rotation matrix

\begin{equation}
\left( 
\begin{array}{ccc}
1 & 0 & 0 \\
0 & \cos{60^{\circ}}  & -\sin{60^{\circ}}  \\
0 & \sin{60^{\circ}}  &  \cos{60^{\circ}} 
\end{array}
\right) 
\end{equation}

\noindent{}to the Cartesian elements (positions and velocities) of the four planets, 
and then transforming back into orbital element space (this transformation does not 
affect $a$, $e$ nor $f$). The result was that the inclination values of the four planets 
ranged from $58.8^{\circ}$ to $60.3^{\circ}$. Because the Sun-like stars were treated
as point masses, they were not affected by, nor affected, this rotation.



\begin{figure*}
\centerline{
\includegraphics[width=9cm]{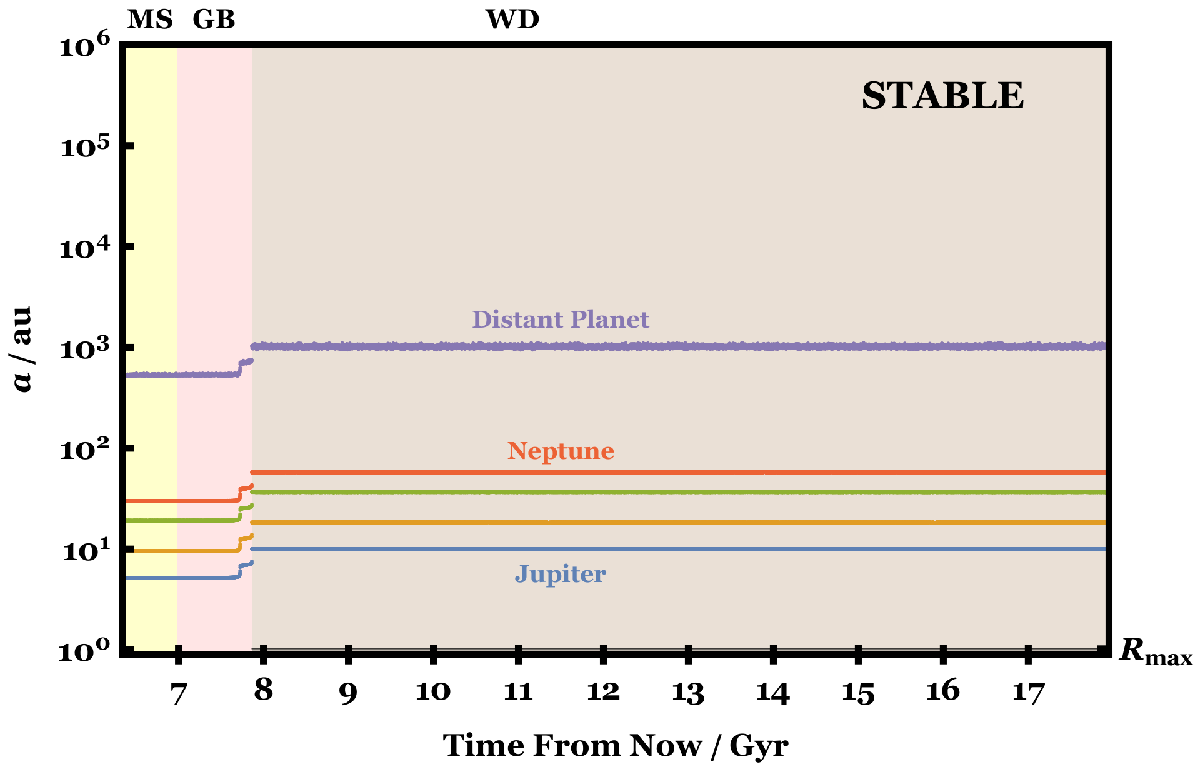}
\includegraphics[width=9cm]{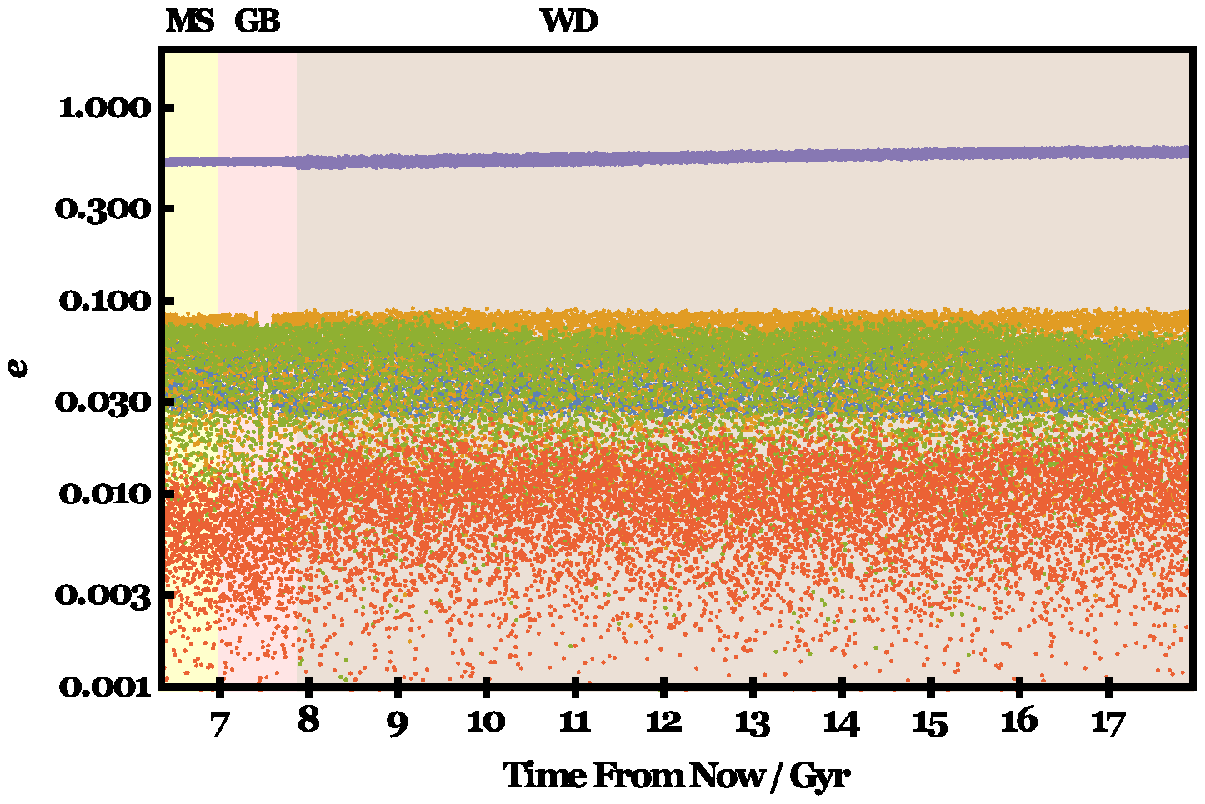}
}
\
\
\centerline{
\includegraphics[width=9cm]{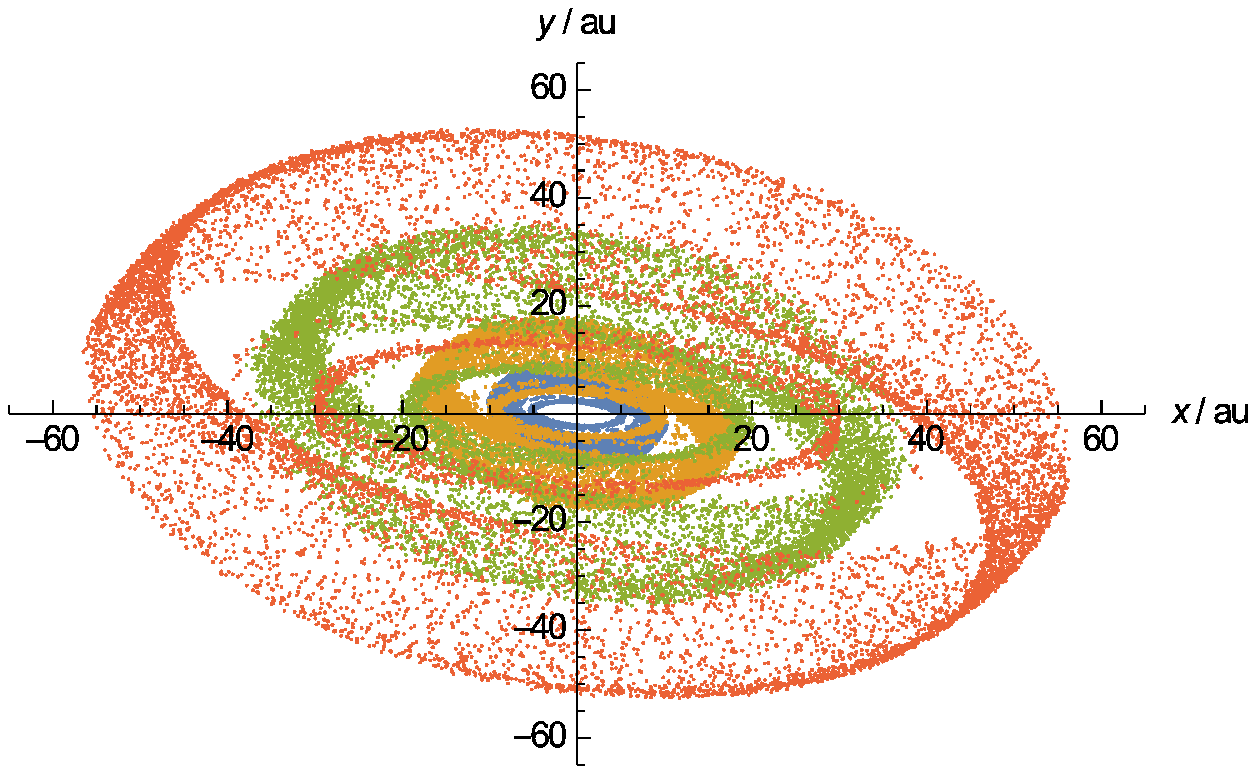}
\includegraphics[width=9cm]{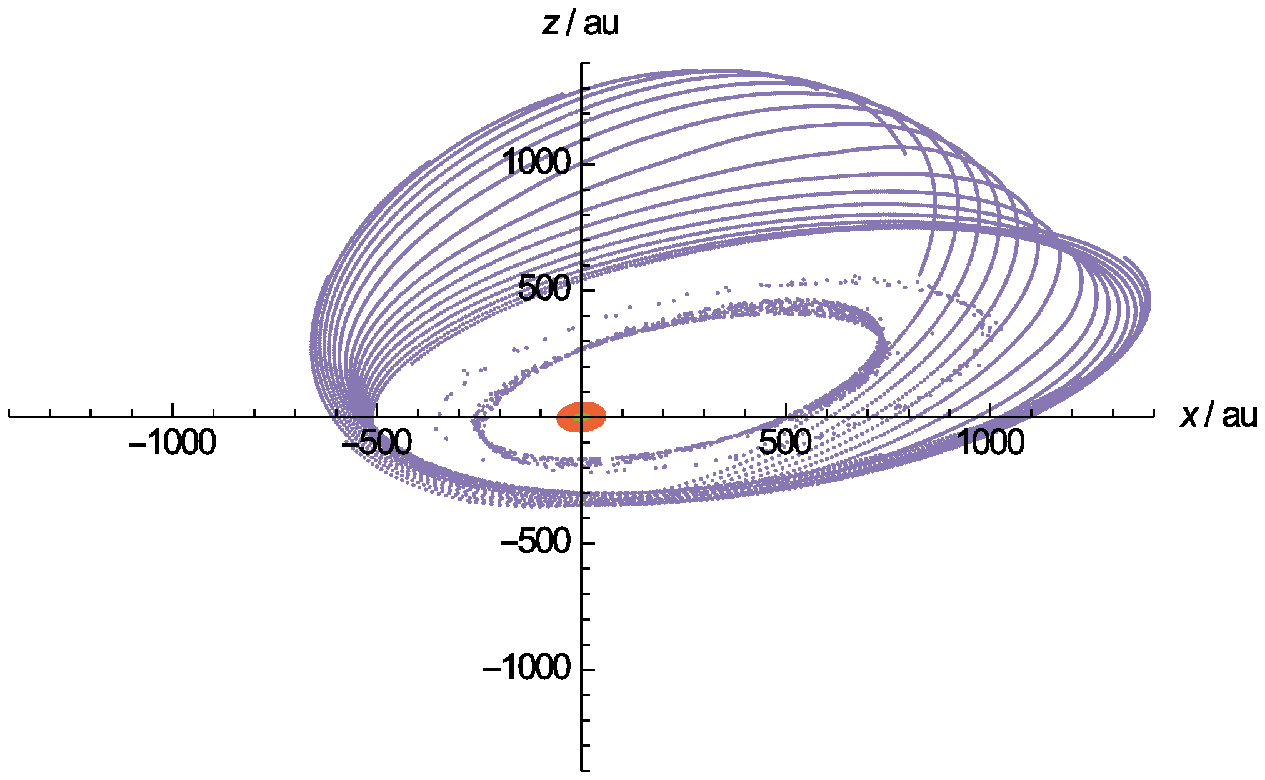}
}
\caption{
A quiet end: one example of an evolutionary
sequence of Jupiter, Saturn, Uranus, Neptune
analogues and a more distant planet with $M_{\rm pl} \approx 82M_{\oplus}$,
$a_{\rm pl}({\rm initial}) \approx 536$ au, $q_{\rm pl}({\rm initial}) \approx 260$ au
and $i_{\rm pl}({\rm initial}) \approx 42.8^{\circ}$ (where $i$ for the other
planets ranges from $58.8^{\circ}$ to $60.2^{\circ}$).
The upper left and right panels illustrate the semimajor
axis and eccentricity evolution of all of the planets during
the end of the main sequence phase (yellow background),
the entire giant branch phase (pink background), and for about
10 Gyr of the white dwarf phase (gray background). $R_{\rm max}$
refers to the maximum Solar radius attained during the giant branch phase.
The bottom panels are also time evolution plots (not snapshots)
but shown on the $x-y$ and $x-z$ planes.
The semimajor axes of all planets are small enough to
ensure adiabatic evolution during mass loss and
an insensitivity to Galactic tides. Consequently, during
mass loss the semimajor axes approximately double and the
eccentricities remain unchanged.
}
\label{eta5_2}
\end{figure*}

\section{Monte-Carlo simulations}

Having laid out the background for my study, I am now ready to present
my Monte-Carlo simulations.

\subsection{Initial conditions}

In order to establish initial conditions, I consider each parameter in 
turn. In all cases, when I use the term
``initial'', that refers to the starting point of my simulations
(10941.0429 Myr from the zero-age-main-sequence)
but also more generally to main sequence values. Time
is measured from the current epoch, which was assumed
to occur exactly 4.6 Gyr after the zero-age-main-sequence.
My adopted ranges for the distant planet's mass, semimajor axis and inclination (below) are all larger than those assumed by recent studies attempting to constrain the properties of a (potential) Planet Nine in the Solar system. These expanded ranges allow me to help identify when instability would occur for planetary systems that generally resemble the Solar system. Naturally if Planet Nine is discovered and its mass, orbit and position are pinpointed, then a dedicated post-main-sequence study which adopts those parameters would be of interest.

\begin{itemize}

\item  I chose to sample $M_{\rm pl}$ randomly from a logarithmic distribution with 
range $10^1 - 10^4 M_{\oplus}$. I assumed that lower masses would not likely be able to
significantly perturb the giant planets even during a close encounter (the masses
of Uranus and Neptune are about 14-17$M_{\oplus}$). Higher masses would
represent stars rather than planets; in fact the typically-assumed
  planet-brown dwarf boundary of 13 Jupiter masses is approximately equal to
  $4100M_{\oplus}$.

\item  I sampled $a_{\rm pl}({\rm initial})$ randomly from a logarithmic distribution
with range $5 \times 10^2 - 5 \times 10^4$~au, and the planet's initial orbital pericentre 
$\equiv q_{\rm pl}({\rm initial})$~$=$~$a_{\rm pl}({\rm initial})
$~$\times\left[1 - e_{\rm pl}({\rm initial}) \right]$
from a uniform distribution with a range of 100 to 400 au. These choices allowed
me to sample the entire system from $10^2$ au out to the edge of the Hill Ellipsoid 
(at $\approx 10^5$ au).


\item I sampled $i_{\rm pl}({\rm initial})$ randomly from a uniform distribution with
range $40^{\circ}-80^{\circ}$, which is centred around the approximately $60^{\circ}$
tilt that I have imposed for the giant planets (recall that $i=0^{\circ}$ corresponds to
the Galactic plane). My range is somewhat arbitrary:
I recognize that a distant planet may have any inclination, particularly if it is a captured
object, but for computational purposes restricted the range to an order of magnitude
greater than the mutual inclinations amongst the giant planets. Sampling non-zero intermediate values of inclination is anyway particularly valuable in order to probe the effect of Galactic tides.

\item The orbital angles $\omega_{\rm pl}({\rm initial})$ and $\Omega_{\rm pl}({\rm initial})$, and
  $\Pi_{\rm pl}({\rm initial})$ were sampled randomly 
from a uniform distribution over their entire ranges.

\end{itemize}

\subsection{Results}

\begin{figure}
\includegraphics[width=9cm]{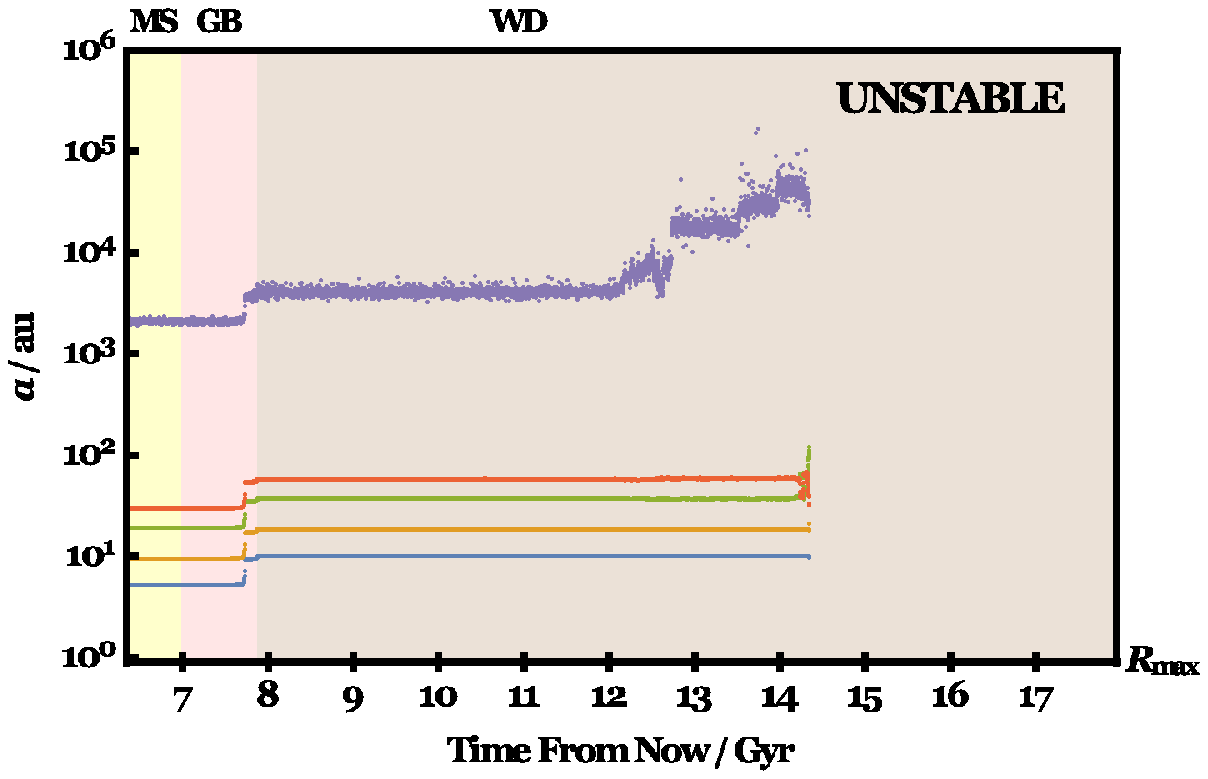}
\includegraphics[width=9cm]{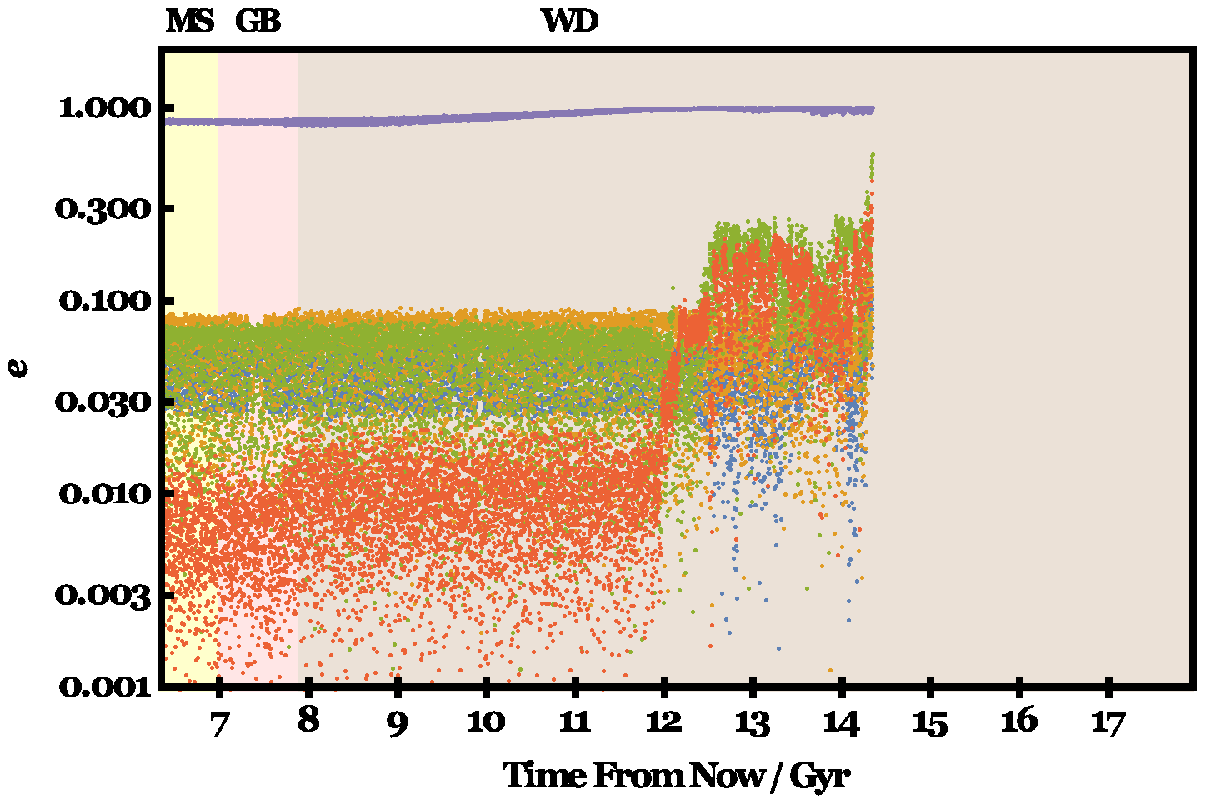}
\includegraphics[width=9cm]{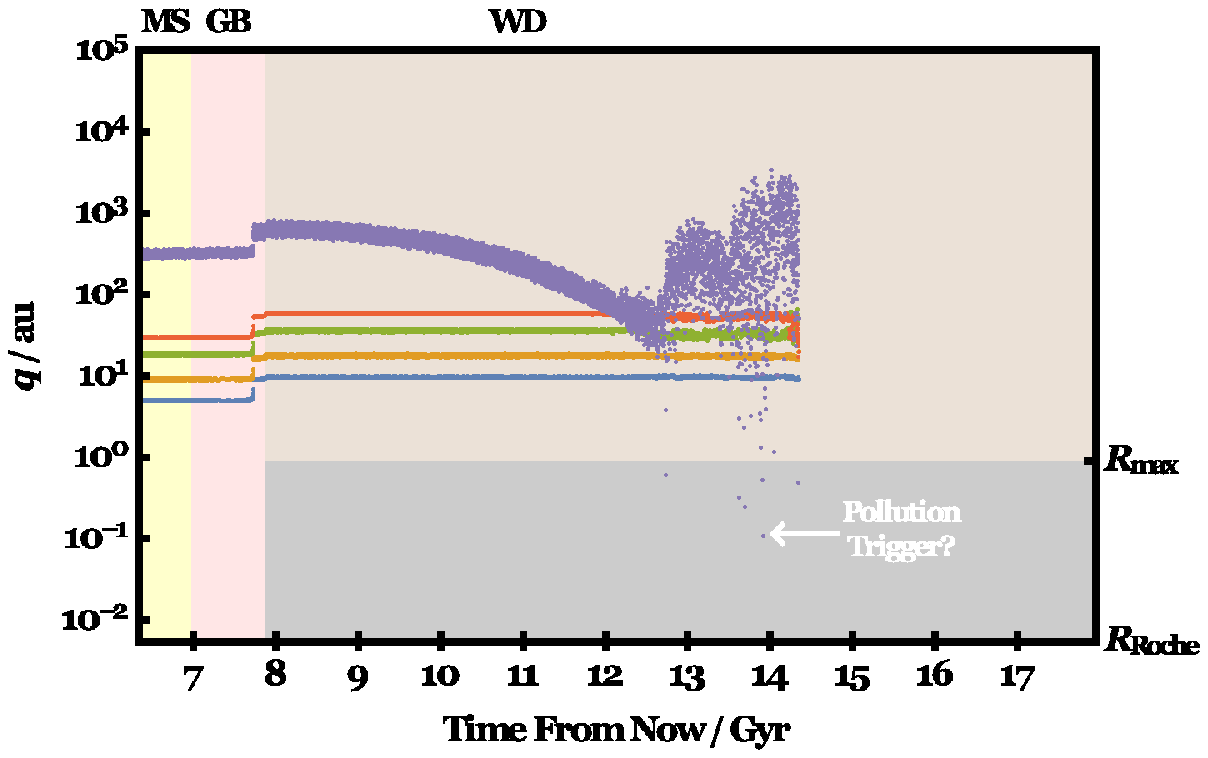}
\caption{
Tidally-induced instability: Here the distant planet
($M_{\rm pl}$~$\approx$~$109M_{\oplus}$) has a large enough initial semimajor axis
$a_{\rm pl}({\rm initial})$~$\approx$~$2030$ au for Galactic
tides to have a noticeable effect during the white
dwarf phase. The tides create an initial increase
in the already high value of $e_{\rm pl}({\rm initial})$~$\approx$~$0.85$,
triggering ejection of the analogue of Neptune as $q_{\rm pl}$ approaches the location
of the other four planets (bottom plot). Eventually,
$q_{\rm pl}$~$<$~$R_{\rm max}$, causing the distant planet to sweep through 
any remaining debris in the inner system, 
which may pollute the eventual white dwarf once inside
its disruption radius $R_{\rm Roche}$.
}
\label{eta8_86}
\end{figure}

\begin{figure}
\includegraphics[width=9cm]{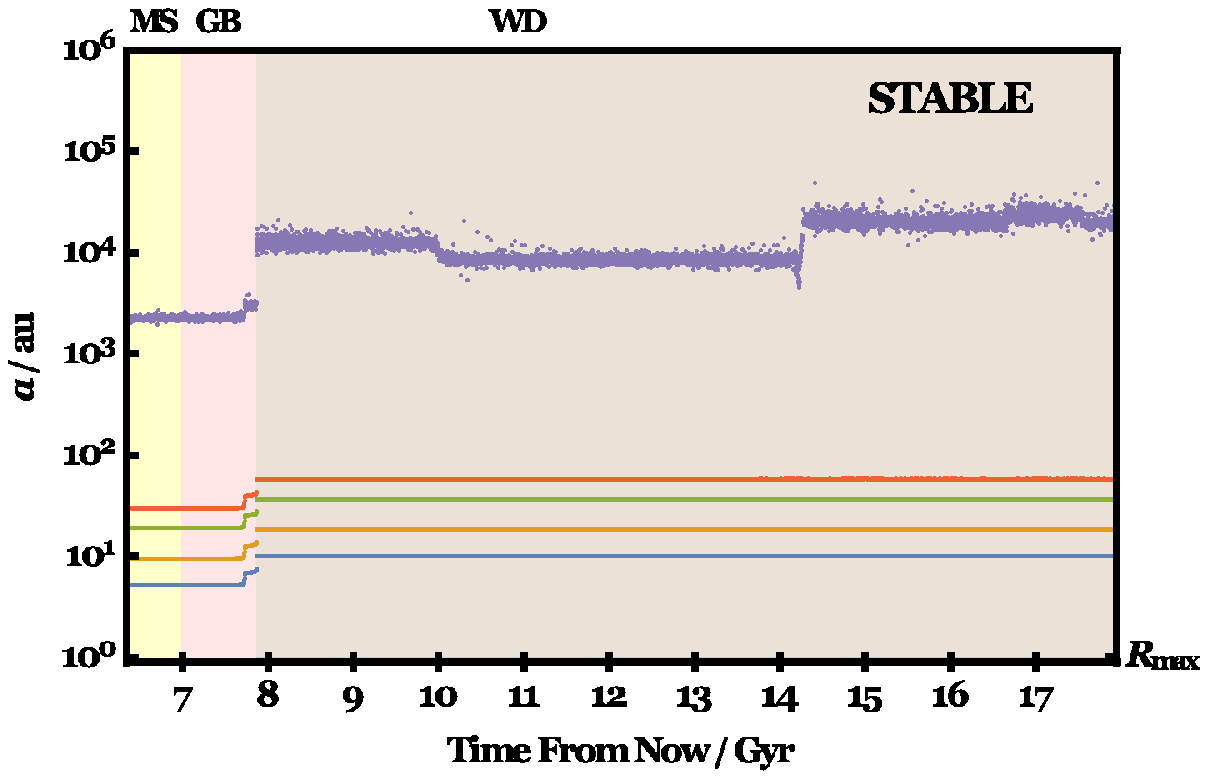}
\includegraphics[width=9cm]{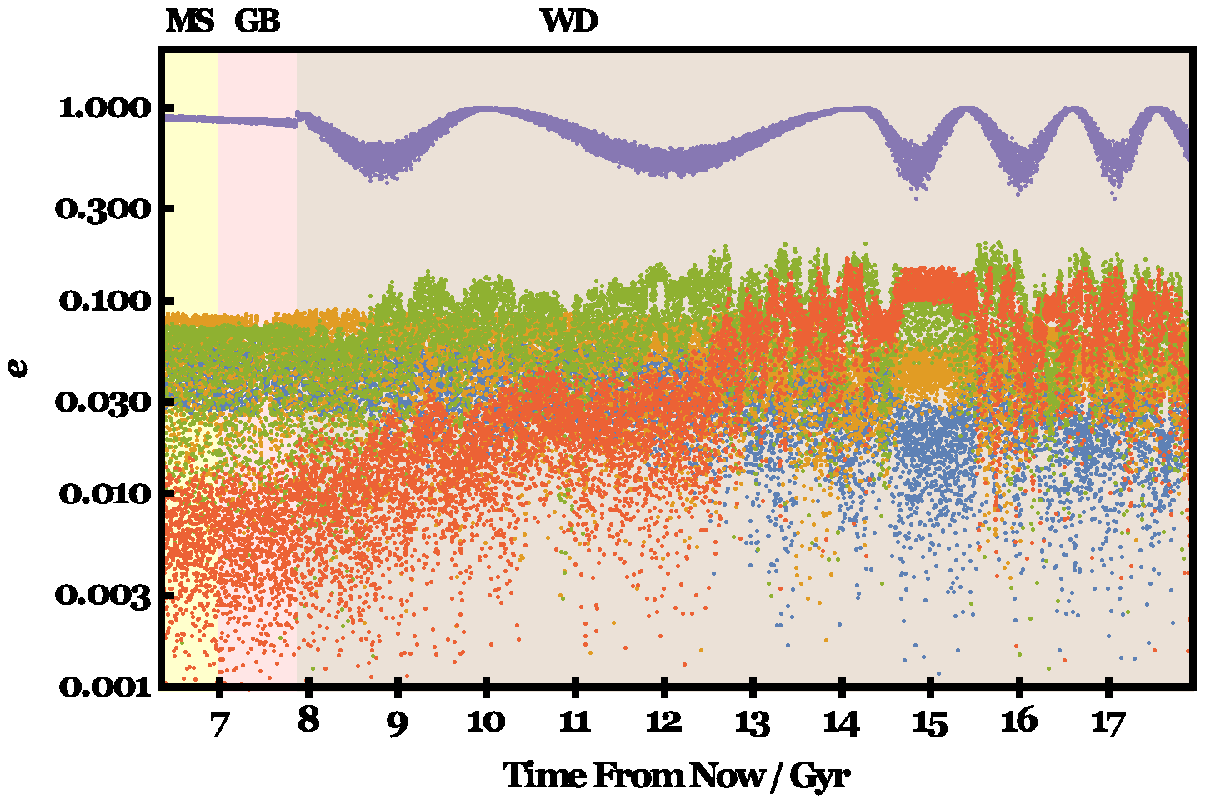}
\includegraphics[width=9cm]{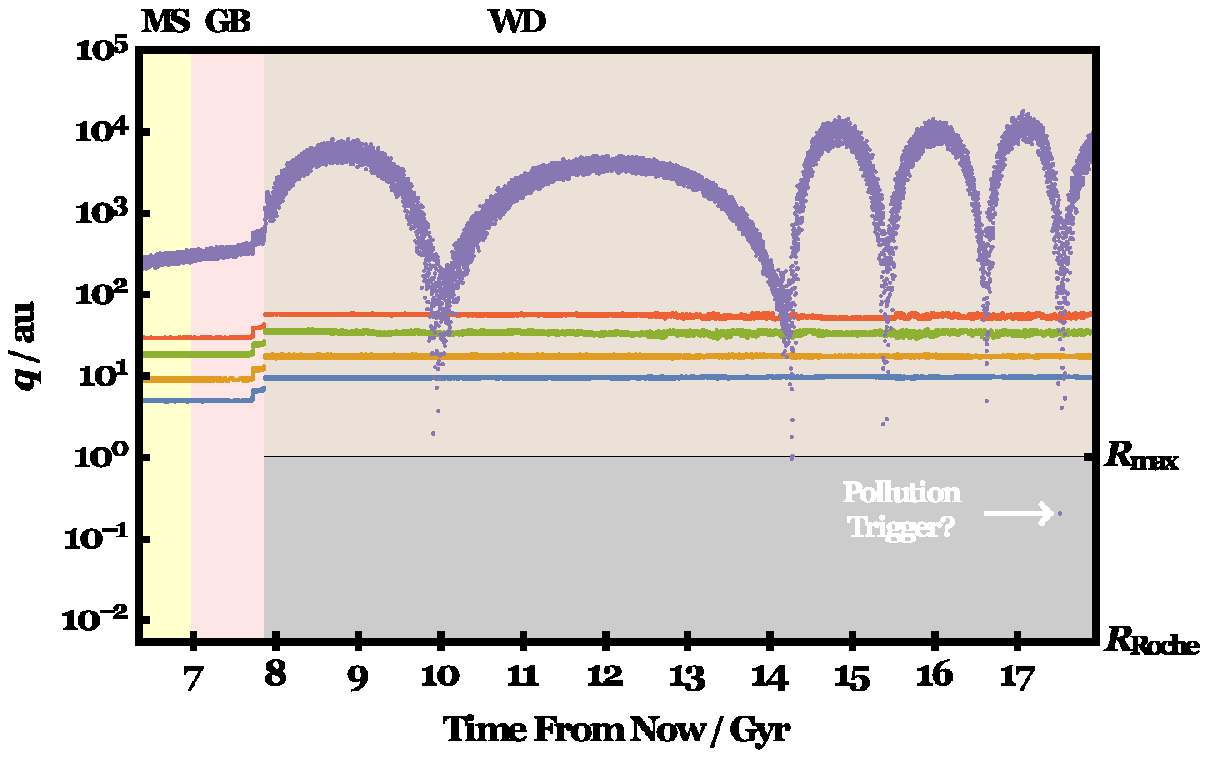}
\caption{
Stability despite tidal excitation: A super-Earth-sized
distant planet ($M_{\rm pl}$~$\approx$~$12M_{\oplus}$)
leaves the main sequence in an orbital 
configuration which causes non-adiabatic evolution,
yielding an order of magnitude increase in 
semimajor axis. Consequently, Galactic tides become
strong, producing oscillations in eccentricity on the
order of tenths. The non-uniformity of the oscillations
and jumps in $a_{\rm pl}$ indicate effects from perturbations with the other
four planets, whose eccentricities remain under 0.2.
The repeated deep radial incursions of the distant planet 
near or within $R_{\rm max}$ might provide pathways
to pollute the eventual white dwarf.
}
\label{eta5_31}
\end{figure}

\begin{figure}
\includegraphics[width=9cm]{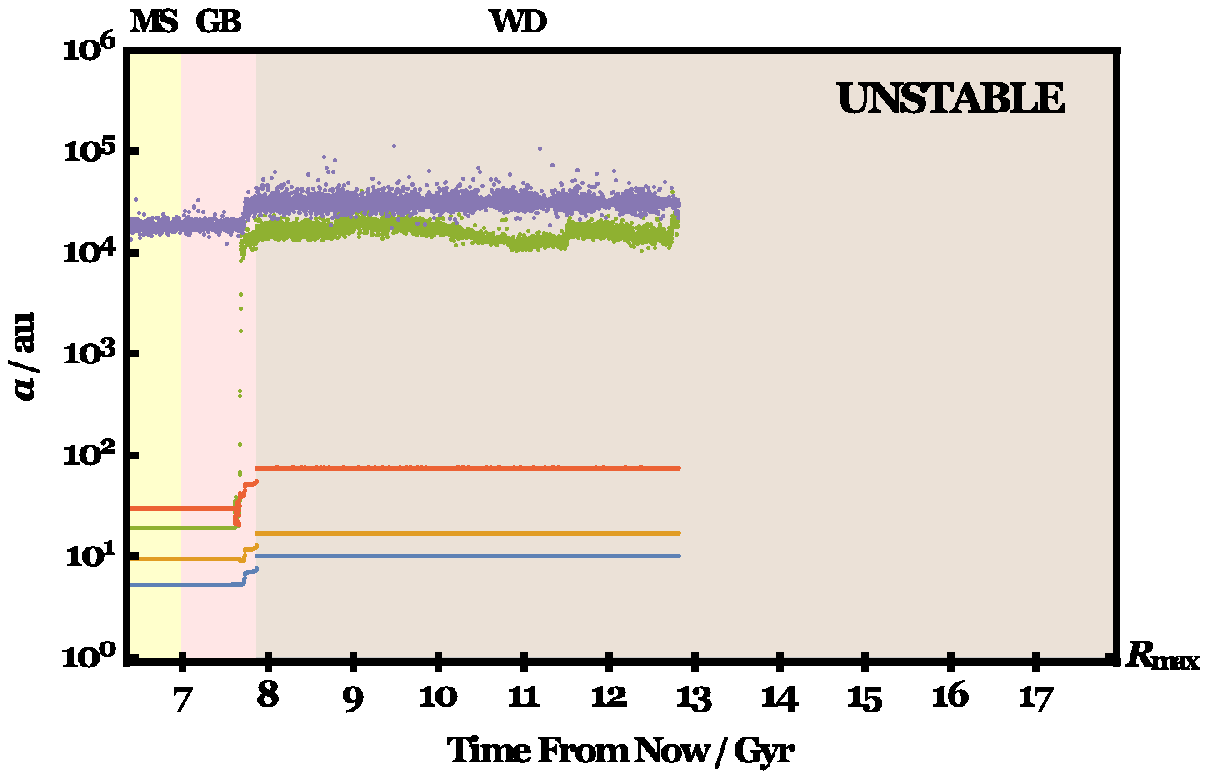}
\includegraphics[width=9cm]{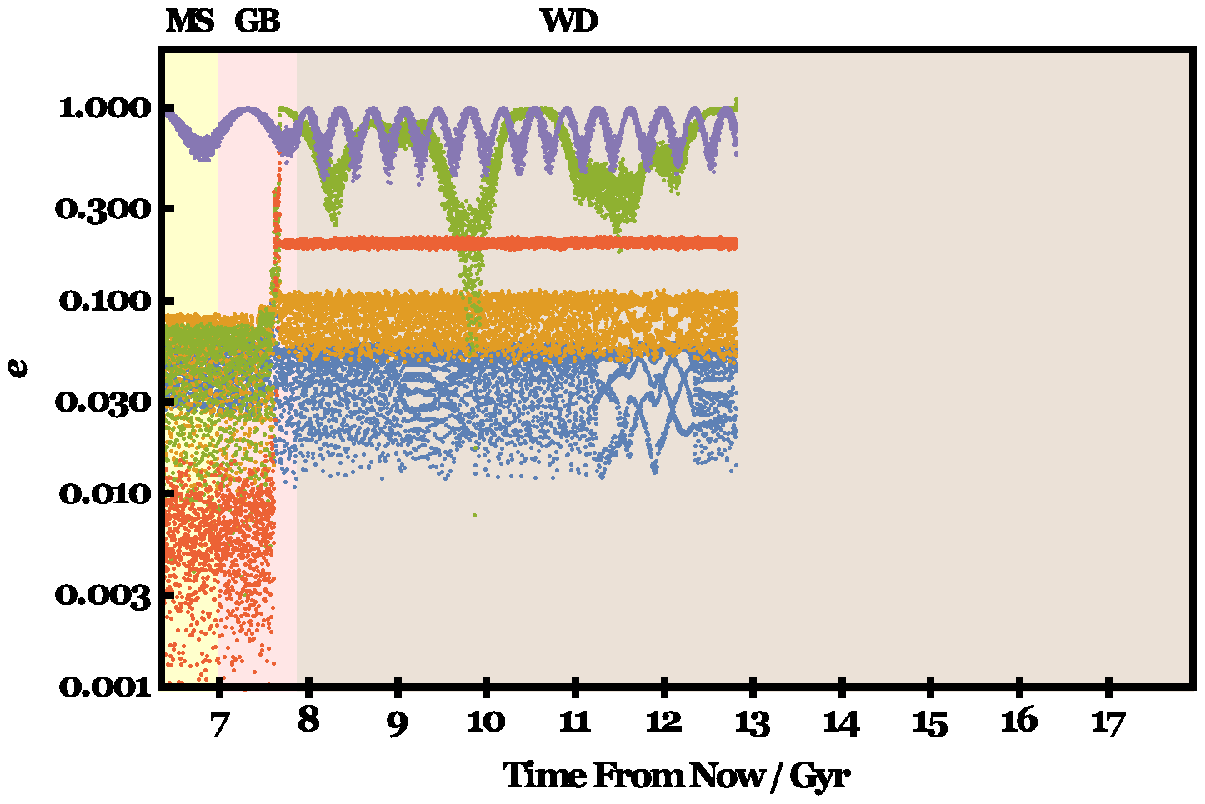}
\includegraphics[width=9cm]{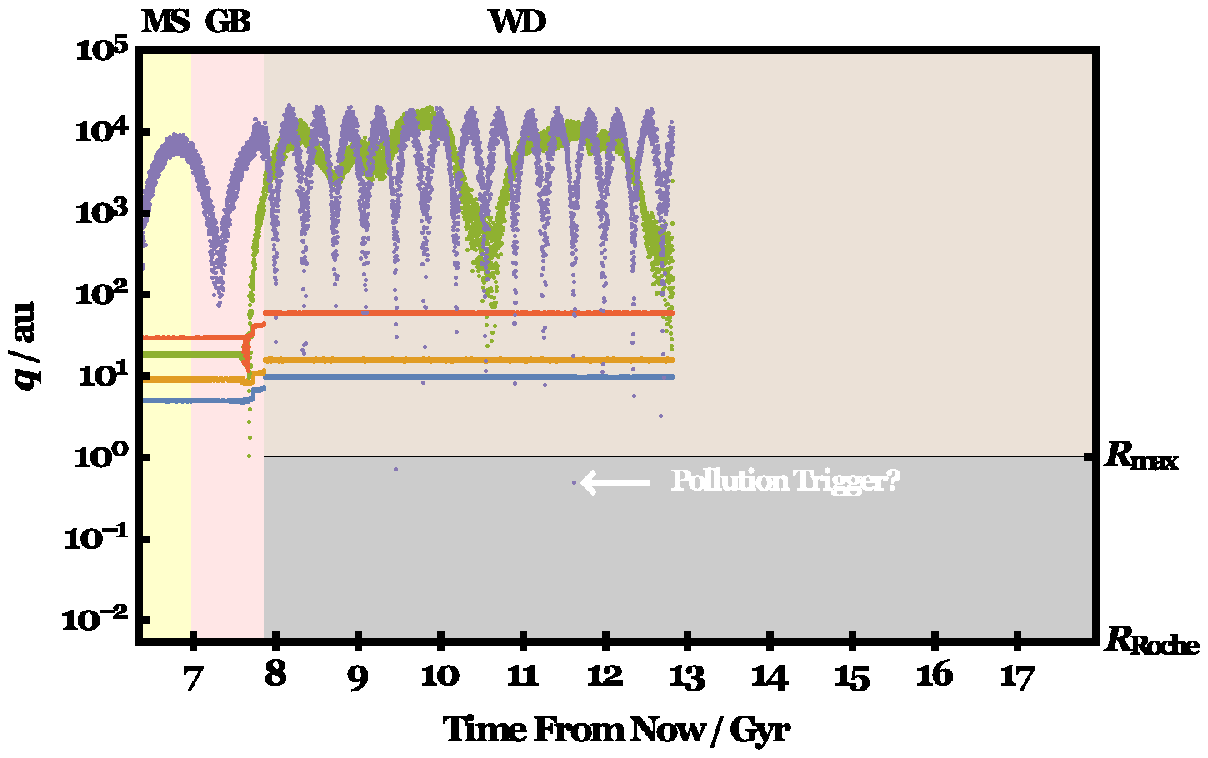}
\caption{
The Uranus-like planet is kidnapped, then ejected:
In this unusual case, mass loss triggers scattering between the Uranus and
Neptune analogues and the distant planet ($M_{\rm pl}$~$\approx$~$806M_{\oplus}$),
which harbours a cometary-like orbit with
$a_{\rm pl}({\rm initial})$~$\approx$~$17000$ au and
$q_{\rm pl}({\rm initial})$~$\approx$~$222$ au.
Because Uranus-analogue's semimajor axis is increased by two orders of 
magnitude, both the Uranus-like planet and the distant planet are then
strongly affected by Galactic tides (middle panel).
Eventually, soon after the Uranus analogue approaches the orbits
of the Neptune, Saturn and Jupiter analogues at a time of about
$13$ Gyr (bottom panel), the Uranus-like planet is ejected. A caveat to this
figure is that stellar tidal effects on the Uranus analogue (not modelled) 
may have played a role during strong scattering on the
star's red giant branch phase.
}
\label{eta5_22}
\end{figure}

\begin{figure}
\includegraphics[width=9cm]{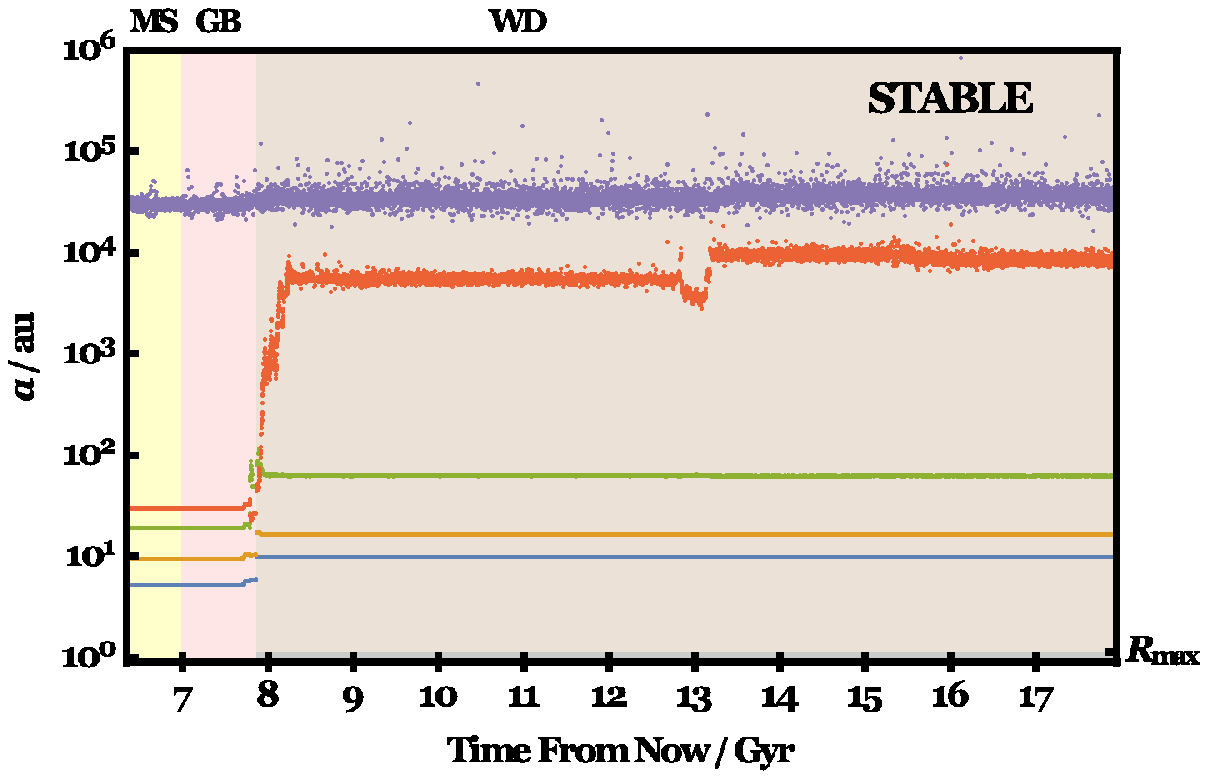}
\includegraphics[width=9cm]{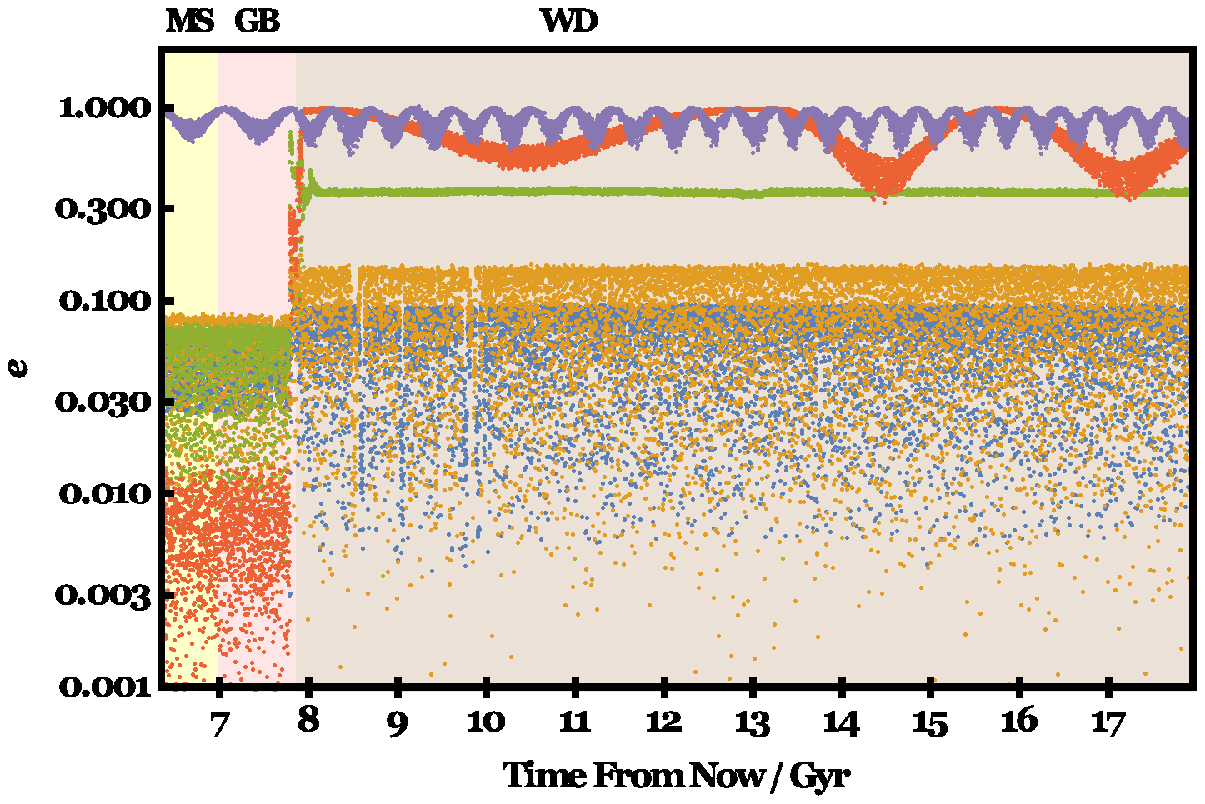}
\includegraphics[width=9cm]{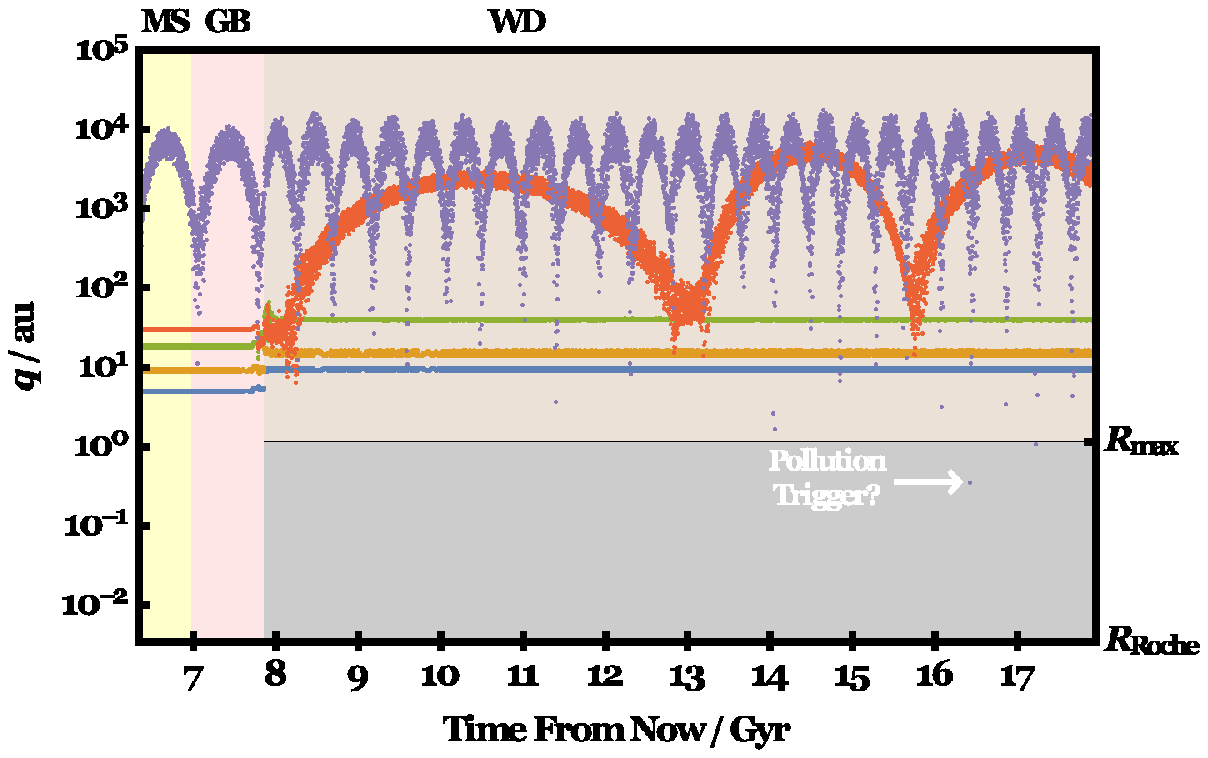}
\caption{
The Neptune-like planet is kidnapped, but the system remains stable:
For this rare case, the tip of the asymptotic
giant branch coincides with a pericentre passage
of the distant planet (bottom panel).  During this time, the
Neptune and Uranus analogues are perturbed enough by the slightly larger
distant planet ($M_{\rm pl}$~$\approx$~$30M_{\oplus}$) to
force Neptune-analogue's semimajor axis to increase by
two orders of magnitude and increase Uranus analogue's eccentricity
to about 0.4. The planet's already non-adiabatic
value of [$a_{\rm pl}({\rm initial})$~$\approx$~$27000$~au] combined
with this scattering event results in a negligible
change in $a_{\rm pl}$. Subsequently, both the Neptune-like and the distant
planet are affected by Galactic tides and sweep through 
the Solar system analogue, 
achieving $q_{\rm pl} < R_{\rm max}$. However, these actions
do not disturb Uranus, Saturn nor Jupiter, and the
system retains all planets until the end of the
simulation.
}
\label{eta2_66}
\end{figure}

I present my results in two stages. The first is
an analysis of specific systems, and the second is
an analysis of the ensemble.

\subsubsection{Specific systems}

I have analyzed the outcomes for each of nearly 300 simulations on an individual basis,
and decided to present here a flavour of the wide variety of outcomes.
In the following figures, I consistently use the same horizontal and vertical
ranges, even for unstable simulations that were terminated.

\begin{itemize}

\item {\bf Figure \ref{eta5_2} (stable)} 
One of the simplest cases was evolved
with a $\eta=0.5$ Sun-like star and features a distant planet
on a short-enough
orbit that neither Galactic tides nor non-adiabatic mass loss ever become factors.
Consequently, the semimajor axis and eccentricity changes were predictable,
and all planets were far away enough from each other such that they
did not suffer close encounters. The eccentricities of the analogues of the four known 
giant planets mirrored their main sequence values and variations, 
whereas the distant planet's eccentricity
was slightly noticeably altered: the amplitude of $e_{\rm pl}$ increased
just after mass loss, and Galactic tides did cause a barely perceptible
secular change over 10 Gyr of white dwarf evolution.

The Cartesian cross-sections (bottom panels) reveal how mass loss changes 
planetary orbits in space.
For each planet, there are two distinct tori of points, corresponding to main
sequence and white dwarf values. For the distant planet, a sparse intermediate ring of points
can be seen, which corresponds to a transition state during the giant branch phase.

\item {\bf Figure \ref{eta8_86} (unstable)} 
Here the distant planet's initial semimajor axis of about 2030 au
was (i) small enough to achieve adiabatic mass loss, (ii) small enough to be
unaffected by Galactic tides on the main sequence and giant branch phases,
(iii) and large enough to be affected by Galactic tides non-negligibly on
the white dwarf phase. Regarding the first point, this semimajor axis lies
within the Solar system adiabaticity transition region of $10^3-10^4$ au,
but remained adiabatic because here $\eta = 0.8$ (see fig. 2 of \citealt*{verwya2012}).   
The modulation of $e_{\rm pl}$ due to Galactic tides was high enough
to be seen on the middle plot. This increase of eccentricity (by about a tenth from 0.85) 
decreased $q_{\rm pl}$ (bottom plot)
gradually, until close encounters pumped up the eccentricities of the Uranus
and Neptune analogues, eventually ejecting the Neptune-like planet
and stopping the simulation.

The value of $q_{\rm pl}$ became much lower than the orbital pericentres of the other planets
before ejection.  Indicated on the right axis of the bottom plot are 
$R_{\rm max}$ and $R_{\rm Roche}$, which represent the maximum radius that the star
attained during the giant branch phase, and the Roche, or disruption radius, of the
white dwarf. For all plots, I computed this radius assuming a Jupiter-mass planet.
In this figure, $q_{\rm pl} < R_{\rm max}$. Consequently, the distant planet swept
through an area of space that had been untouched since the end of the giant branch phase. Hence, 
extant debris from the inner system (such as e.g. planets like Mars, 
and remnants of asteroid belt-like breakups; Veras et al. 2014a), could be perturbed
towards the white dwarf and pollute it. This same behaviour was observed
in \cite{vergae2015} and \cite{veretal2016a}.

\item {\bf Figure \ref{eta5_31} (stable)}
Here the distant planet's initial semimajor axis of 2230 au is very similar
to that in Fig. \ref{eta8_86}, except in this instance the mass loss evolution
was non-adiabatic. The reason partly is, for this simulation, $\eta = 0.5$
(see fig. 2 of \citealt*{verwya2012}). The resulting large jump 
in the value of $a_{\rm pl}$, by nearly a factor of 10, created a situation where
Galactic tides played a large role during white dwarf evolution.

Whereas Galactic tides slightly decreased $e_{\rm pl}$ on the main sequence
and white dwarf phases, the large jump in $a_{\rm pl}$ created discernable
oscillations during the white dwarf phase. If Jupiter, Saturn, Uranus
and Neptune analogues were not present, then these oscillations would have had a constant
amplitude and frequency. However, mutual interactions triggered semimajor
axis ``jumps'' (upper plot) plus a corresponding change in shape of the
eccentricity oscillations. Further, over the first 7 Gyr of white dwarf evolution,
Neptune-analogue's eccentricity increased by over an order of magnitude, to about 0.1-0.2.
Nevertheless, over the duration of the simulation, the system remained stable.

The bottom plot illustrates deep radial incursions (within or close to
$R_{\rm max}$) of the distant planet towards the 
white dwarf: one at 10 Gyr, and then four more from 14-18 Gyr. These
illustrate how a dynamically-active inner planetary system environment
can be triggered at any white dwarf age, potentially explaining white
dwarf pollution at different epochs \citep{bonver2015,hampor2016,petmun2016}.

\begin{figure*}
\centerline{
\includegraphics[width=9cm]{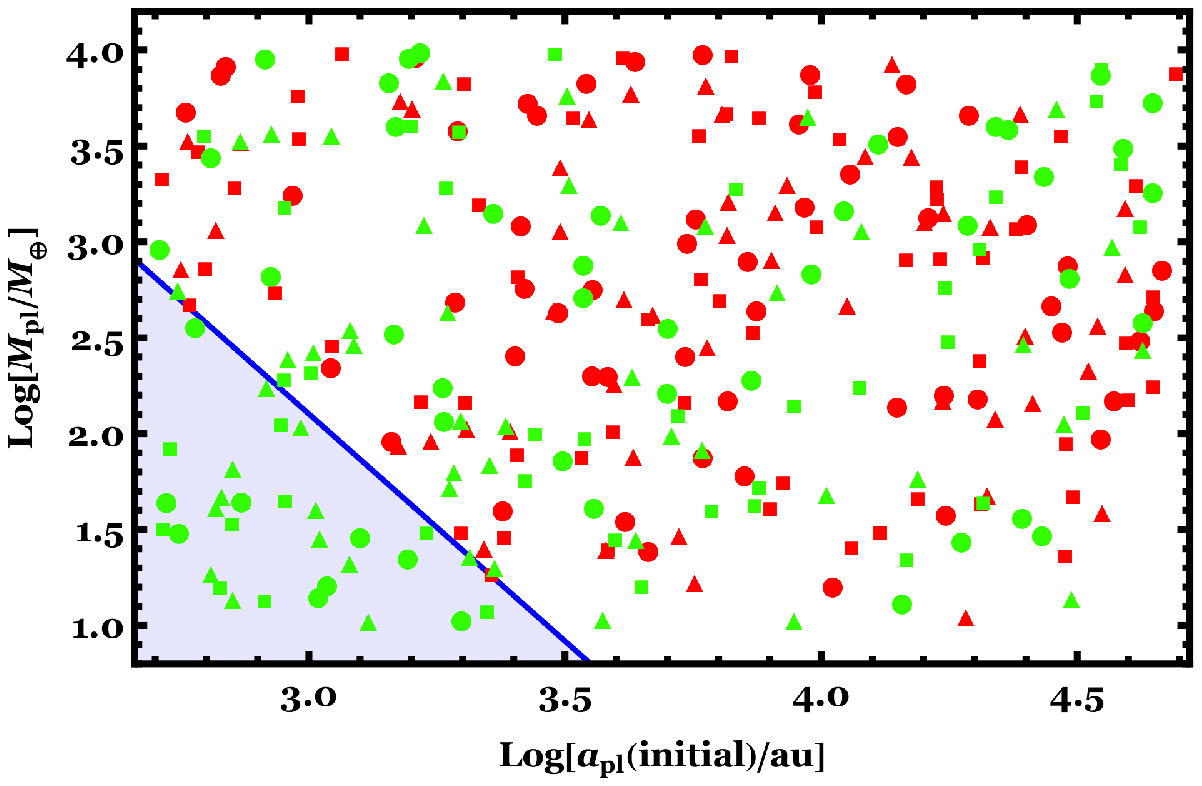}
\includegraphics[width=9cm]{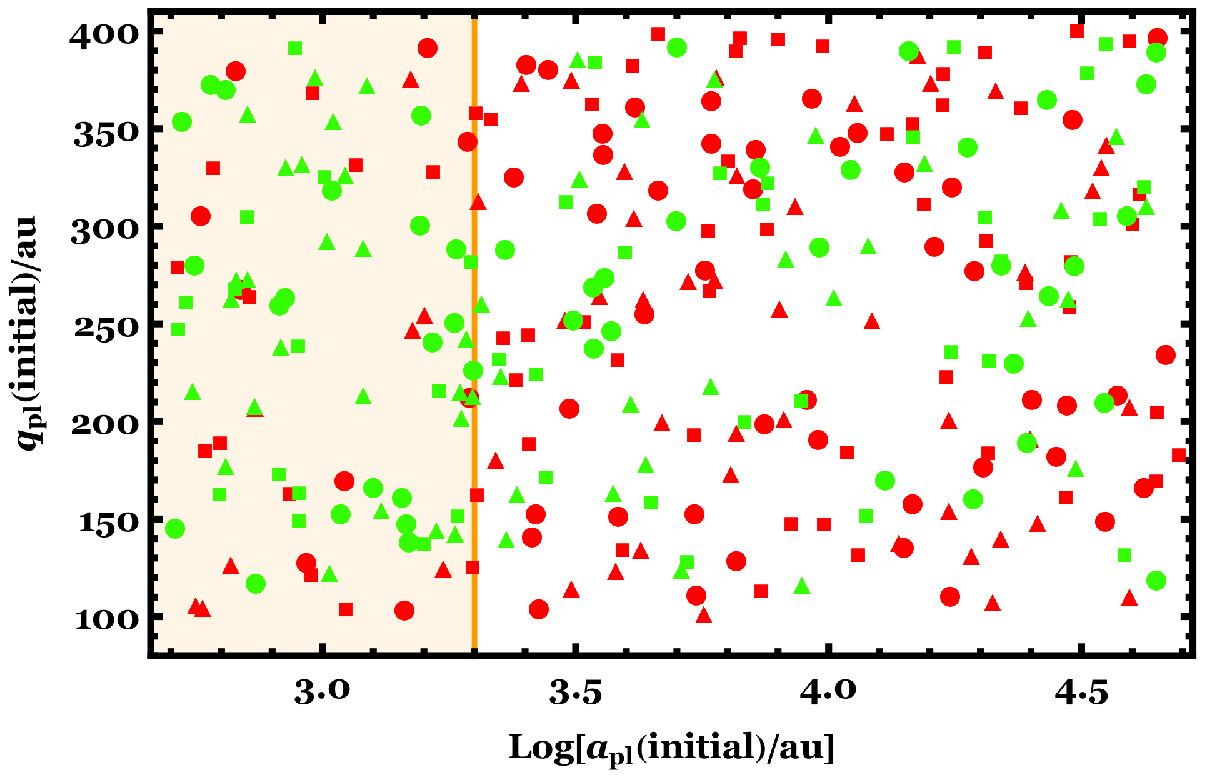}
}
\
\
\centerline{
\includegraphics[width=9cm]{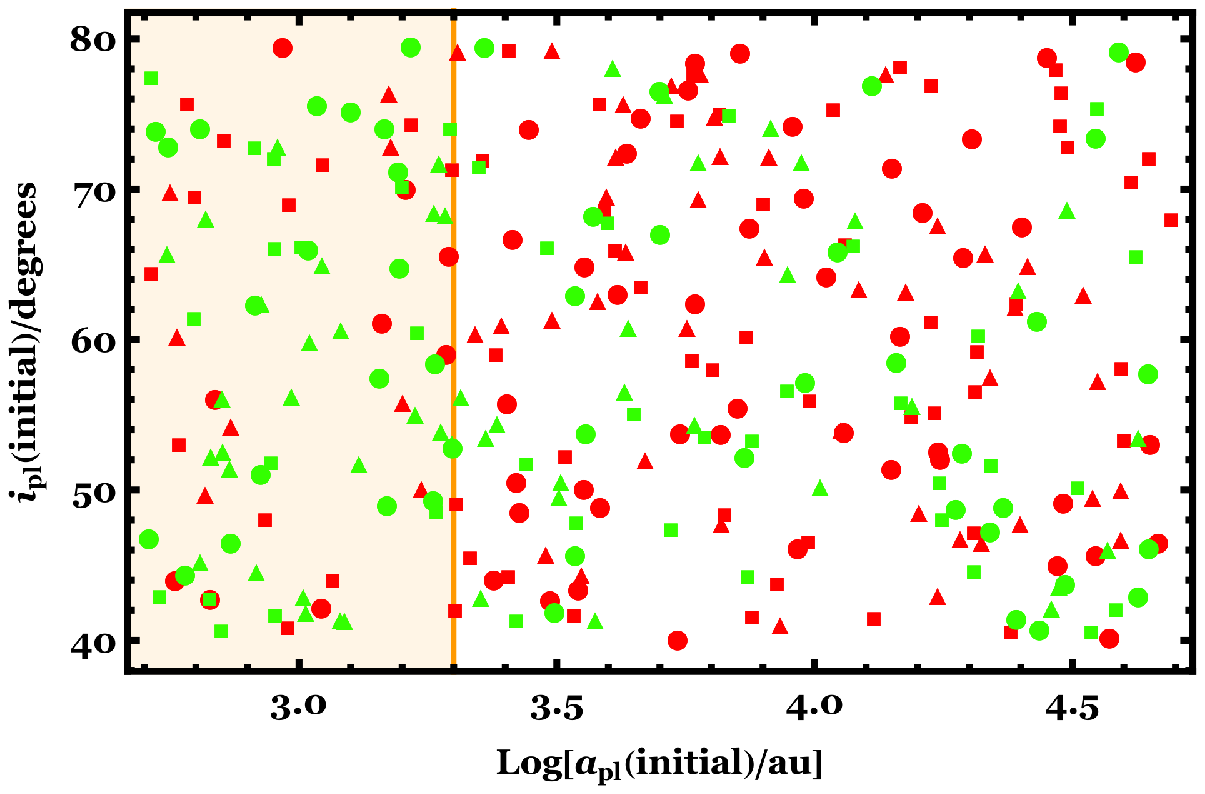}
\includegraphics[width=9cm]{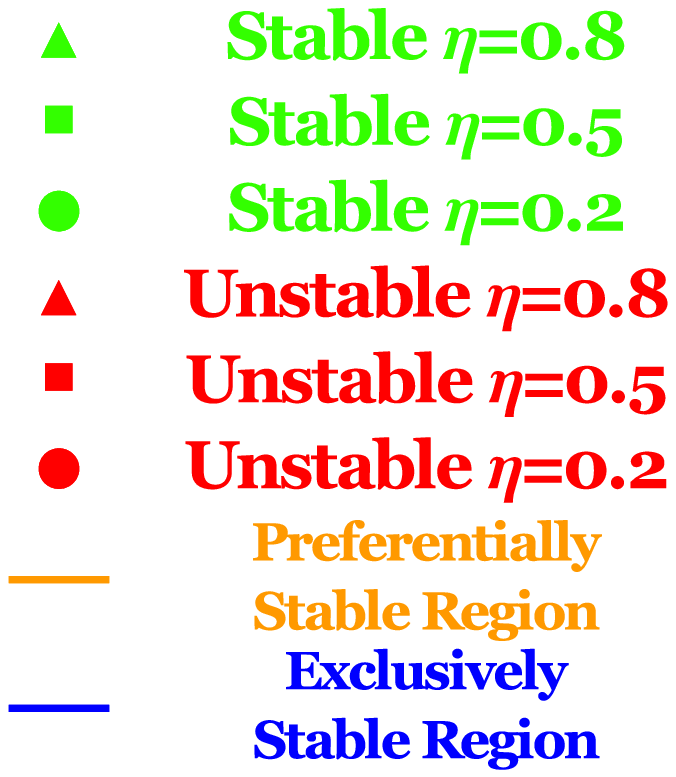}
}
\caption{
The outcomes from all simulations. Stable simulations are indicated
in green symbols, and unstable ones are in red.  The different shapes
refer to different Solar models; these plots illustrate that the results
are independent of $\eta$. Both stable and unstable simulations
populate all areas of all plots, except for the blue triangle in
the upper-left plot; this triangle contains all stable simulations.
Orange rectangles indicate regions where stability is more likely
to occur than not.
The outcomes are largely independent of $q_{\rm pl}({\rm initial})$
and $i_{\rm pl}({\rm initial})$.
}
\label{main}
\end{figure*}

\item {\bf Figure \ref{eta5_22} (unstable)}
A highly eccentric ($\gtrsim 0.9$) distant planet with $a_{\rm pl} > 10^4$ au
is likely to generate interesting dynamical behaviour because
it is sure to evolve non-adiabatically from mass loss and always
is affected significantly by Galactic tides. For this system,
these attributes provided an environment which was conducive to strong
scattering events, allowing the analogue of Uranus to be ``captured'' into a 
$10^4$ au orbit. This orbit existed for an appreciable 5 Gyr -- longer
than the Solar system's current age -- before the analogue of Uranus was ejected.

Some other noteworthy dynamical aspects of this $\eta = 0.5$ system are:
(i) the change in frequency of the distant planet's eccentricity 
oscillations from the giant branch to white dwarf phases (due to
semimajor axis increase), (ii) the Uranus-like planet achieving an orbit with
a pericentre of nearly 1 au at the end of the
red giant branch phase (but before the asymptotic giant branch phase); this
fact places a significant caveat on this figure because tidal
interactions with the star (not modelled) could have retarded the planet's
evolution during this stage \citep{schcon2008}, (iii) Neptune, Jupiter and Saturn 
analogues settling into orbits whose secular eccentricity evolution was well-behaved,
compared to that from any other figures.

\item {\bf Figure \ref{eta2_66} (stable)}
In this figure, $a_{\rm pl}({\rm initial})$~$\approx$~$27000$~au, a value 59 per cent
greater than the corresponding value in Fig. \ref{eta5_22}, and the initial
eccentricities in both figures coincidentally differed by just $0.05$ per cent. Despite
these values, the amplitude of the eccentricity oscillations due to Galactic tides
in this figure was not as high. The reason is because of the distant planet's
other orbital parameters:
$i_{\rm pl}$, $\omega_{\rm pl}$, and $\Omega_{\rm pl}$ all played a significant, non-trivial role.

Regardless, the pericentre passages were deep enough to coincide with the orbits of
the four giant planet analogues. One of these passages occurred close to the tip of the
asymptotic giant branch (whereas in Fig. \ref{eta5_22} a passage occurred close to the tip
of the red giant branch). The resulting scattering event helped contribute to
$a_{\rm pl}$'s net negligible change at the giant branch / white dwarf boundary, and triggered
a semimajor axis
increase for the analogue of Neptune to almost $10^4$ au after a few hundred Myr. Neptune-analogue's
eccentricity then became strongly affected by Galactic tides. Subsequently, both the Neptune-like planet
and the distant planet experienced repeated radial incursions
-- at different rates -- providing
a dynamically active and ever-changing environment potentially conducive to pollution.
The Uranus, Saturn and Jupiter analogues were not bothered by this activity, and the system
remained stable for the duration of the simulation.

\end{itemize}

\subsubsection{System ensemble}

The primary goal of this study is to determine the stability boundary: for what
mass and orbital parameters of a distant planet will instability occur. Amongst the different
potential combinations of variables to plot, I found that the most revealing combinations
are those seen in Fig. \ref{main}. The figure reveals the following points.

\begin{itemize}

\item Any potentially sharp boundary is limited to the blue triangle in the upper-left plot;
in this triangle every system sampled was stable.
Even so, just a few tens of simulations within that triangle were performed. Otherwise, 
unstable and stable simulations appear on all regions of all plots, showcasing how sensitive
the simulation outcome is to all orbital parameters, including $\Omega_{\rm pl}$, $\omega_{\rm pl}$ and $\Pi_{\rm pl}$.

The blue triangle makes sense: Although the notion that smaller values of $a_{\rm pl}$ would yield
greater stability might seem counterintuitive, in fact such values guarantee that the mass loss 
evolution is adiabatic and that Galactic tides play no significant role in the dynamics. Also,
the greater the value of $M_{\rm pl}$, the more likely that the distant planet could
scatter off of the other planets and create instability.

\item Mass aside, stability is more generally likely to occur than not when
  $a_{\rm pl}({\rm initial})$ is within a few thousand au, as indicated by the orange
  regions. The reason is the same as above. 

\item There is no discernable stability dependence on $\eta$ (for values between 0.2 and 0.8).

\item There is no discernable stability dependence on $q_{\rm pl}$ (for values between 100 and 400 au).

\item There is no discernable stability dependence on $i_{\rm pl}$ (for values within $20^{\circ}$ below and above the ecliptic).

\end{itemize}

Other trends from the ensemble of plots include details of the instability: 97
per cent of all unstable simulations
featured ejections. The remainder were engulfments (defined to occur when a planet
intersected the star's Roche radius).  In no case did the Jupiter-like planet become unstable.
The analogues of Saturn, Uranus and Neptune and the distant planet 
were either ejected or engulfed 8, 53, 24 and 15 per cent of the time, respectively.
This result makes sense given that in almost every simulation, the analogue of Uranus was the
least massive planet. \cite{veretal2016a}
showed that scattering amongst unequal-mass planets across all phases of stellar evolution will
preferentially eject the least massive planet (just as along the main sequence alone).

\section{Conclusions}

I demonstrated that a distant planet with an orbital pericentre under 400 au
could pose a serious danger to the stability of Solar system analogues
during a Sun-like star's giant branch and white dwarf phases. This statement
holds true for a distant planet which is at least as massive as
Jupiter and harbours a semimajor axis beyond about 300 au, or for a super-Earth
when its semimajor axis exceeds about 3000 au. The driver for the
instability is a combination of Galactic tides and stellar mass
loss, which together or separately may induce close encounters amongst the
five planets, with the distant planet always representing the trigger.

These results have implications for both the Solar system and for extrasolar systems. The existence of a trans-Neptunian planet could eventually eliminate (likely through ejection) at least one of the giant planets (most likely Uranus, then Neptune) and rearrange the others, but only if this ``extra'' planet is massive enough, distant enough and its orbit is appropriately oriented with respect to the Galactic plane and the existing giants. This planet may represent the purported Planet Nine, a hidden other planet, or a planet that will be captured later during the Sun's main sequence evolution. Because the Sun will become a white dwarf and contain strewn-about debris from a destroyed asteroid belt (Veras et al. 2014a) plus a charred Mars and perhaps some liberated moons \citep{payetal2016a,payetal2016b}, an ample reservoir of extant material in the inner Solar system would be available to be perturbed into the Solar white dwarf, ``polluting'' it.

The consequences for other planetary systems are profound. Multiple planets beyond about 5 au (such as analogues of Jupiter, Saturn, Uranus and Neptune) may be common, but are so far unfortunately effectively hidden from detection by Doppler radial velocity and transit photometry techniques, the two most successful planet-finding techniques.  If more distant, trans-Neptunian-like planets are also common, then the ingredients may exist to regularly generate instability and a frequently-changing dynamical environment during white dwarf phases of evolution. Such movement could provide a natural way to perturb inner system debris into white dwarfs at a variety of white dwarf ages and rates, helping to explain current observations.

\section*{Acknowledgements}

I thank the referee for a particularly thorough and precise consideration of the manuscript, and Konstantin Batygin for his perspectives and insightful comments on an earlier version. I have received funding from the European Research Council under the European Union's Seventh Framework Programme (FP/2007-2013)/ERC Grant Agreement n. 320964 (WDTracer).




\label{lastpage}

\begin{thebibliography}{99}

\bibitem[Adams \& Bloch(2013)]{adablo2013} Adams F.~C., Bloch A.~M.\ 2013, ApJL, 777, L30 

\bibitem[Alcock, Fristrom \& Siegelman(1986)]{alcetal1986for3} Alcock C., Fristrom C.~C., Siegelman R.\ 1986, ApJ, 302, 462
  
\bibitem[Alonso et al.(2016)]{aloetal2016} Alonso R., Rappaport S., Deeg H.~J., Palle E.\ 2016, A\&A, 589, L6 

\bibitem[Barber et al.(2016)]{baretal2016} Barber S.~D., Belardi C., Kilic M., Gianninas A.\ 2016, MNRAS, 459, 1415 

\bibitem[Barnes \& Greenberg(2006)]{bargre2006} Barnes R., Greenberg R.\ 2006, ApJL, 647, L163 

\bibitem[Barnes \& Greenberg(2007)]{bargre2007} Barnes R., Greenberg R.\ 2007, ApJL, 665, L67 

\bibitem[Batygin \& Brown(2016)]{batbro2016} Batygin K., Brown, M.~E.\ 2016, AJ, 151, 22 
  
\bibitem[Batygin \& Laughlin(2008)]{batlau2008} Batygin K., Laughlin G.\ 2008, ApJ, 683, 1207-1216 

\bibitem[Batygin, Morbidelli \& Holman(2015)]{batetal2015for3} Batygin K., Morbidelli A., Holman M.~J.\ 2015, ApJ, 799, 120 
  
\bibitem[Bergfors et al.(2014)]{beretal2014} Bergfors C., Farihi J., Dufour P., Rocchetto M.\ 2014, MNRAS, 444, 2147 

\bibitem[Beust(2016)]{beust2016} Beust H.\ 2016, A\&A, 590, L2 

\bibitem[Bochkarev \& Rafikov(2011)]{bocraf2011} Bochkarev K.~V., Rafikov R.~R.\ 2011, ApJ, 741, 36 


\bibitem[Bonsor \& Veras(2015)]{bonver2015} Bonsor A., Veras D.\ 2015, MNRAS, 454, 53

\bibitem[Bonsor \& Wyatt(2012)]{bonwya2012} Bonsor A., Wyatt M.~C.\ 2012, MNRAS, 420, 2990

\bibitem[Bonsor, Mustill \& Wyatt(2011)]{bonetal2011for3} Bonsor A., Mustill A.~J., Wyatt M.~C.\ 2011, MNRAS, 414, 930 


  
\bibitem[Brasser(2001)]{Br01} Brasser R.\ 2001, MNRAS, 324, 1109 

\bibitem[Breiter \& Ratajczak(2005)]{Br05} Breiter S., Ratajczak R.\ 2005, MNRAS, 364, 1222

\bibitem[Breiter, Dybczynski \& Elipe(1996)]{Br96for3} Breiter S., Dybczynski P.~A., Elipe A.\ 1996, A\&A, 315, 618 
  
  
\bibitem[Bromley \& Kenyon(2016)]{brokey2016} Bromley B.~C., Kenyon S.~J.\ 2016, ApJ, 826, 64 


\bibitem[Brown \& Batygin(2016)]{brobat2016} Brown M.~E., Batygin K.\ 2016, ApJ, 824, L23 

\bibitem[Brown, Trujillo \& Rabinowitz(2004)]{broetal2004for3} Brown M.~E., Trujillo C., Rabinowitz D.\ 2004, ApJ, 617, 645

\bibitem[Campante, Barclay \& Swift(2015)]{cametal2015for3} Campante T.~L., Barclay T., Swift J.~J., et al.\ 2015, ApJ, 799, 170 


\bibitem[Chambers(1999)]{chambers1999} Chambers J.~E.\ 1999, MNRAS, 304, 793

\bibitem[Chambers, Wetherill \& Boss(1996)]{chaetal1996for3} Chambers J.~E., Wetherill G.~W., Boss A.~P.\ 1996, Icarus, 119, 261


\bibitem[Chatterjee et al.(2008)]{chaetal2008} Chatterjee S., Ford E.~B., Matsumura S., Rasio F.~A.\ 2008, ApJ, 686, 580-602 

\bibitem[Cowan, Holder \& Kaib(2016)]{cowetal2016for3} Cowan N.~B., Holder G., Kaib N.~A.\ 2016, ApJL, 822, L2 
  
\bibitem[Croll et al.(2016)]{croetal2016} Croll B., Dalba P.~A., Vanderburg A., et al.\ 2016, Submitted to ApJL, arXiv:1510.06434 

\bibitem[Davies et al.(2014)]{davetal2014} Davies M.~B., Adams F.~C., Armitage P., et al.\ 2014, Protostars and Planets VI, 787 

\bibitem[de la Fuente Marcos \& de la Fuente Marcos(2014)]{deldel2014} de la Fuente Marcos C., de la Fuente Marcos R.\ 2014, MNRAS, 443, L59 

\bibitem[de la Fuente Marcos \& de la Fuente Marcos(2016a)]{deldel2016a} de la Fuente Marcos C., de la Fuente Marcos R.\ 2016a, MNRAS, 460, L64

\bibitem[de la Fuente Marcos \& de la Fuente Marcos(2016b)]{deldel2016b} de la Fuente Marcos C., de la Fuente Marcos R.\ 2016b, MNRAS, 459, L66

\bibitem[de la Fuente Marcos, de la Fuente Marcos \& Aarseth(2016)]{deletal2016for3} de la Fuente Marcos C., de la Fuente Marcos R., Aarseth S.~J.\ 2016, MNRAS, 460, L123
  
\bibitem[Debes \& Sigurdsson(2002)]{debsig2002} Debes J.~H., Sigurdsson S.\ 2002, ApJ, 572, 556 

\bibitem[Debes, Walsh \& Stark(2012)]{debetal2012for3} Debes J.~H., Walsh K.~J., Stark C.\ 2012, ApJ, 747, 148

\bibitem[Deck, Payne \& Holman(2013)]{decetal2013for3} Deck K.~M., Payne M., Holman M.~J.\ 2013, ApJ, 774, 129

\bibitem[Dosopoulou \& Kalogera(2016a)]{doskal2016a} Dosopoulou F., Kalogera V.\ 2016a, ApJ In Press, arXiv:1603.06592 

\bibitem[Dosopoulou \& Kalogera(2016b)]{doskal2016b} Dosopoulou F., Kalogera V.\ 2016b, ApJ In Press, arXiv:1603.06593 

\bibitem[Dufour et al.(2007)]{dufetal2007} Dufour P., Bergeron P., Liebert J., et al.\ 2007, ApJ, 663, 1291

\bibitem[Duncan \& Lissauer(1998)]{dunlis1998} Duncan M.~J., Lissauer J.~J.\ 1998, Icarus, 134, 303 

\bibitem[Farihi(2016)]{farihi2016} Farihi J.\ 2016, New Astronomy Reviews, 71, 9 

\bibitem[Farihi, Jura \& Zuckerman(2009)]{faretal2009for3} Farihi J., Jura M., Zuckerman B.\ 2009, ApJ, 694, 805

\bibitem[Farihi, G{\"a}nsicke \& Koester(2013)]{faretal2013} Farihi J., G{\"a}nsicke B.~T. Koester D.\ 2013, Science, 342, 218 
  
\bibitem[Fienga et al.(2016)]{fieetal2016} Fienga A., Laskar J., Manche H., Gastineau M.\ 2016, A\&A, 587, L8 

\bibitem[Fortney et al.(2016)]{foretal2016} Fortney J.~J. et al.\ 2016, ApJL, 824, L25 

\bibitem[Fouchard(2004)]{Fo04} Fouchard M.\ 2004, MNRAS, 349, 347  
  
\bibitem[Fouchard et al.(2006)]{Fo06} Fouchard, M., Froeschl{\'e}, C., Valsecchi, G., \& Rickman, H.\ 2006, Celestial Mechanics and Dynamical Astronomy, 95, 299  

\bibitem[Frewen \& Hansen(2014)]{frehan2014} Frewen S.~F.~N., Hansen B.~M.~S.\ 2014, MNRAS, 439, 2442 

\bibitem[G{\"a}nsicke et al.(2006)]{gaeetal2006} G{\"a}nsicke B.~T., Marsh T.~R., Southworth J., Rebassa-Mansergas A.\ 2006, Science, 314, 1908 

\bibitem[G{\"a}nsicke et al.(2016)]{ganetal2016} G{\"a}nsicke B.~T., Aungwerojwit A., Marsh T.~R., et al.\ 2016, ApJL, 818, L7 

\bibitem[Ginzburg, Sari \& Loeb(2016)]{ginetal2016for3} Ginzburg S., Sari R., Loeb A.\ 2016, ApJL, 822, L11
  
\bibitem[Gladman \& Chan(2006)]{glacha2006} Gladman B., Chan C.\ 2006, ApJL, 643, L135 

\bibitem[Gomes, Soares \& Brasser(2015)]{gometal2015for3} Gomes R.~S., Soares J.~S., Brasser R.\ 2015, Icarus, 258, 37

\bibitem[Gurri, Veras \& G{\"a}nsicke(2016)]{guretal2016for3} Gurri, P. Veras, D. G{\"a}nsicke B.~T.\ 2016, Submitted to MNRAS
  
\bibitem[Hadjidemetriou(1963)]{hadjidemetriou1963} Hadjidemetriou J.~D.\ 1963, Icarus, 2, 440 

\bibitem[Hamers \& Portegies Zwart(2016)]{hampor2016} Hamers A.~S., Portegies Zwart S.~F.\ 2016, MNRAS In Press, arXiv:1607.01397 

\bibitem[Heisler \& Tremaine(1986)]{He86} Heisler J., Tremaine S.\ 1986, Icarus, 65, 13 

\bibitem[Holman \& Payne(2016a)]{holpay2016a} Holman M.~J., Payne M.~J.\ 2016a, AJ, in press (arXiv:1603.09008)

\bibitem[Holman \& Payne(2016b)]{holpay2016b} Holman, M.~J., \& Payne, M.~J.\ 2016b, AAS journals, submitted (arXiv:1604.03180)

\bibitem[Hurley, Pols \& Tout(2000)]{huretal2000for3} Hurley J.~R., Pols O.~R., Tout C.~A.\ 2000, MNRAS, 315, 543 
  
\bibitem[Iorio(2012)]{iorio2012} Iorio L.\ 2012, Celestial Mechanics and Dynamical Astronomy, 112, 117 

\bibitem[Iorio(2014)]{iorio2014} Iorio L.\ 2014, MNRAS, 444, L78 

\bibitem[Iorio(2015)]{iorio2015} Iorio L.\ 2015, arXiv:1512.05288 

\bibitem[J{\'{\i}}lkov{\'a} et al.(2015)]{jiletal2015} J{\'{\i}}lkov{\'a} L., Portegies Zwart S., Pijloo T., Hammer M.\ 2015, MNRAS, 453, 3157 

\bibitem[Jura \& Xu(2010)]{jurxu2010} Jura M., Xu S.\ 2010, AJ, 140, 1129 

\bibitem[Jura \& Xu(2012)]{jurxu2012} Jura M., Xu S.\ 2012, AJ, 143, 6 
  
\bibitem[Kenyon \& Bromley(2016)]{kenbro2016} Kenyon S.~J., Bromley B.~C.\ 2016, ApJ, 825, 33

\bibitem[Kepler et al.(2015)]{kepetal2015} Kepler S.~O., Pelisoli I., Koester D., et al.\ 2015, MNRAS, 446, 4078 

\bibitem[Kepler et al.(2016)]{kepetal2016} Kepler S.~O., Pelisoli I., Koester D., et al.\ 2016, MNRAS, 455, 3413

\bibitem[Kleinman et al.(2013)]{kleetal2013} Kleinman S.~J., Kepler S.~O., Koester D., et al.\ 2013, ApJS, 204, 5

\bibitem[Koester, G{\"a}nsicke \& Farihi(2014)]{koeetal2014for3} Koester D., G{\"a}nsicke B.~T., Farihi J.\ 2014, A\&A, 566, A34
  
\bibitem[Kratter \& Perets(2012)]{kraper2012} Kratter K.~M., Perets H.~B.\ 2012, ApJ, 753, 91 

\bibitem[Kunitomo et al.(2011)]{kunetal2011} Kunitomo M., Ikoma M., Sato B., Katsuta Y., Ida S.\ 2011, ApJ, 737, 66 

\bibitem[Lagadec \& Zijlstra(2008)]{lagzij2008} Lagadec E., Zijlstra A.~A.\ 2008, MNRAS, 390, L59 

\bibitem[Laskar \& Gastineau(2009)]{lasgas2009} Laskar J., Gastineau M.\ 2009, Nature, 459, 817 

\bibitem[Lawler et al.(2016)]{lawetal2016} Lawler S.~M., Shankman C., Kaib N., et al.\ 2016, Submitted to ApJ, arXiv:1605.06575 

\bibitem[Li \& Adams(2016)]{liada2016} Li G., Adams F.~C.\ 2016, ApJL, 823, L3 

\bibitem[Linder \& Mordasini(2016)]{linmor2016} Linder E.~F., Mordasini C.\ 2016, A\&A, 589, A134 

\bibitem[Luhman(2014)]{luhman2014} Luhman K.~L.\ 2014, ApJ, 781, 4 

\bibitem[Malamud \& Perets(2016)]{malper2016} Malamud U., Perets H.~B.\ 2016, arXiv:1608.00593 

\bibitem[Malhotra, Volk \& Wang(2016)]{maletal2016for3} Malhotra R., Volk K., Wang X.\ 2016, ApJ, 824, L22

\bibitem[Mamajek et al.(2015)]{mametal2015} Mamajek E.~E., Barenfeld S.~A., Ivanov V.~D., et al.\ 2015, ApJL, 800, L17 

\bibitem[Manser et al.(2016)]{manetal2016} Manser C.~J., G{\"a}nsicke B.~T., Marsh T.~R., et al.\ 2016, MNRAS, 455, 4467 

\bibitem[Marzari(2014)]{marzari2014} Marzari F.\ 2014, MNRAS, 442, 1110 

\bibitem[Matese \& Whitman(1989)]{Ma89} Matese J.~J., Whitman P.~G.\ 1989, Icarus, 82, 389

\bibitem[Matese \& Whitman(1992)]{Ma92} Matese J.~J., Whitman P.~G.\ 1992, Celestial Mechanics and Dynamical Astronomy, 54, 13  

\bibitem[Matese et al.(1995)]{Ma95} Matese J.~J., Whitman P.~G., Innanen K.~A., Valtonen M.~J.\ 1995, Icarus, 116, 255

\bibitem[Metzger, Rafikov \& Bochkarev(2012)]{metetal2012for3} Metzger B.~D., Rafikov R.~R., Bochkarev K.~V.\ 2012, MNRAS, 423, 505
  
\bibitem[Murray \& Holman(1997)]{murhol1997} Murray N., Holman M.\ 1997, AJ, 114, 1246 

\bibitem[Mustill \& Villaver(2012)]{musvil2012} Mustill A.~J., Villaver E.\ 2012, ApJ, 761, 121 

\bibitem[Mustill, Marshall \& Villaver(2013)]{musetal2013for3} Mustill A.~J., Marshall J.~P., Villaver E., et al.\ 2013, MNRAS, 436, 2515

\bibitem[Mustill, Veras \& Villaver(2014)]{musetal2014for3} Mustill A.~J., Veras D., Villaver E.\ 2014, MNRAS, 437, 1404

\bibitem[Mustill, Raymond \& Davies(2016)]{musetal2016for3} Mustill A.~J., Raymond S.~N., Davies M.~B.\ 2016, MNRAS, 460, L109

\bibitem[Nakajima, Morino \& Fukagawa(2010)]{naketal2010for3} Nakajima T., Morino J.-I., Fukagawa M.\ 2010, AJ, 140, 713 
  
\bibitem[Nordhaus \& Spiegel(2013)]{norspi2013} Nordhaus J., Spiegel D.~S.\ 2013, MNRAS, 432, 500 

\bibitem[North et al.(2016)]{noretal2016} North T.S.~H. et al. 2016, Submitted to MNRAS

\bibitem[Omarov(1962)]{omarov1962} Omarov T.~B. 1962, Izv. Astrofiz. Inst. Acad. Nauk. KazSSR, 14, 66

\bibitem[Parriott \& Alcock(1998)]{paralc1998} Parriott J., Alcock C.\ 1998, ApJ, 501, 357 

\bibitem[Payne et al.(2016a)]{payetal2016a} Payne M.~J., Veras D., Holman M.~J., G\"{a}nsicke B.~T.\ 2016a, MNRAS, 457, 217 

\bibitem[Payne et al.(2016b)]{payetal2016b} Payne M.~J., Veras D., G\"{a}nsicke B.~T., Holman M.~J. \ 2016b, Submitted to MNRAS

\bibitem[Perets \& Kouwenhoven(2012)]{perkou2012} Perets H.~B., Kouwenhoven M.~B.~N.\ 2012, ApJ, 750, 83 

\bibitem[Petrovich(2015)]{petrovich2015} Petrovich C.\ 2015, ApJ, 808, 120 

\bibitem[Petrovich \& Mu{\~n}oz(2016)]{petmun2016} Petrovich C., Mu{\~n}oz D.~J.\ 2016, Submitted to ApJ, arXiv:1607.04891 

\bibitem[Portegies Zwart(2013)]{portegieszwart2013} Portegies Zwart S.\ 2013, MNRAS, 429, L45 

\bibitem[Pu \& Wu(2015)]{puwu2015} Pu B., Wu Y.\ 2015, ApJ, 807, 44 

\bibitem[Raddi et al.(2015)]{radetal2015} Raddi R., G{\"a}nsicke B.~T., Koester D., et al.\ 2015, MNRAS, 450, 2083 
  
\bibitem[Rafikov(2011a)]{rafikov2011a} Rafikov R.~R.\ 2011a, MNRAS, 416, L55 

\bibitem[Rafikov(2011b)]{rafikov2011b} Rafikov R.~R.\ 2011b, ApJL, 732, L3 

\bibitem[Rafikov \& Garmilla(2012)]{rafgar2012} Rafikov R.~R., Garmilla J.~A.\ 2012, ApJ, 760, 123
 
\bibitem[Rappaport et al.(2016)]{rapetal2016} Rappaport S., Gary B.~L., Kaye T., et al.\ 2016, MNRAS, 458, 3904 

\bibitem[Reimers(1975)]{reimers1975} Reimers D.\ 1975, Memoires of the Societe Royale des Sciences de Liege, 8, 369 

\bibitem[Reimers(1977)]{reimers1977} Reimers D.\ 1977, A\&A, 61, 217 

\bibitem[Ro{\v s}kar et al.(2008)]{rosetal2008} Ro{\v s}kar R., Debattista V.~P., Quinn T.~R., Stinson G.~S., Wadsley J.\ 2008, ApJL, 684, L79 

\bibitem[Rybicki \& Denis(2001)]{rybden2001} Rybicki K.~R., Denis C.\ 2001, Icarus, 151, 130 

\bibitem[Schr{\"o}der \& Connon Smith(2008)]{schcon2008} Schr{\"o}der K.-P., Connon Smith R.\ 2008, MNRAS, 386, 155 

\bibitem[Schr{\"o}der \& Cuntz(2005)]{schcun2005} Schr{\"o}der K.-P., Cuntz M.\ 2005, ApJL, 630, L73 

\bibitem[Sellwood \& Binney(2002)]{selbin2002} Sellwood J.~A., Binney J.~J.\ 2002, MNRAS, 336, 785 

\bibitem[Silva Aguirre et al.(2015)]{siletal2015} Silva Aguirre V., Davies G.~R., Basu S., et al.\ 2015, MNRAS, 452, 2127 

\bibitem[Spiegel \& Madhusudhan(2012)]{spimad2012} Spiegel D.~S., Madhusudhan N.\ 2012, ApJ, 756, 132 

\bibitem[Staff et al.(2016)]{staetal2016} Staff J.~E., De Marco O., Wood P., Galaviz P., Passy J.-C.\ 2016, MNRAS, 458, 832 

\bibitem[Stone, Metzger \& Loeb(2015)]{stoetal2015for3} Stone N., Metzger B.~D., Loeb A.\ 2015, MNRAS, 448, 188 
  
\bibitem[Sumi et al.(2011)]{sumetal2011} Sumi T., Kamiya K., Bennett D.~P., et al.\ 2011, Nature, 473, 349 

\bibitem[Toth(2016)]{toth2016} Toth I. 2016, A\&A, 592, A86
  
\bibitem[Trujillo \& Sheppard(2014)]{trushe2014} Trujillo C.~A., Sheppard S.~S.\ 2014, Nature, 507, 471 

\bibitem[Vanderburg et al.(2015)]{vanetal2015} Vanderburg A., Johnson J.~A., Rappaport S., et al.\ 2015, Nature, 526, 546 

\bibitem[Varvoglis, Sgardeli \& Tsiganis(2012)]{varetal2012for3} Varvoglis H., Sgardeli V., Tsiganis K.\ 2012, Celestial Mechanics and Dynamical Astronomy, 113, 387 
  
\bibitem[Vassiliadis \& Wood(1993)]{vaswoo1993} Vassiliadis E., Wood P.~R.\ 1993, ApJ, 413, 641 








  
\bibitem[Veras(2016)]{veras2016} Veras D.\ 2016, Royal Society Open Science, 3, 150571

\bibitem[Veras \& Evans(2013a)]{vereva2013a} Veras D., Evans N.~W.\ 2013a, MNRAS, 430, 403 

\bibitem[Veras \& Evans(2013b)]{vereva2013b} Veras D., Evans N.~W.\ 2013b, CeMDA, 115, 123
  
\bibitem[Veras \& G\"{a}nsicke(2015)]{vergae2015} Veras D., G\"{a}nsicke B.~T.\ 2015, MNRAS, 447, 1049 

\bibitem[Veras \& Moeckel(2012)]{vermoe2012} Veras D., Moeckel N.\ 2012, MNRAS, 425, 680
  
\bibitem[Veras \& Mustill(2013)]{vermus2013} Veras D., Mustill A.~J.\ 2013, MNRAS, 434, L11

\bibitem[Veras \& Wyatt(2012)]{verwya2012} Veras D., Wyatt M.~C.\ 2012, MNRAS, 421, 2969 

\bibitem[Veras et al.(2011)]{veretal2011} Veras D., Wyatt M.~C., Mustill A.~J., Bonsor A., Eldridge J.~J.\ 2011, MNRAS, 417, 2104 

\bibitem[Veras et al.(2013a)]{veretal2013a} Veras D., Mustill A.~J., Bonsor A., Wyatt M.~C.\ 2013a, MNRAS, 431, 1686 

\bibitem[Veras, Hadjidemtriou \& Tout(2013b)]{veretal2013bfor3} Veras D., Hadjidemetriou J.~D., Tout C.~A.\ 2013b, MNRAS, 435, 2416

\bibitem[Veras, Jacobson \& G\"{a}nsicke(2014a)]{veretal2014afor3} Veras D., Jacobson S.~A., G\"{a}nsicke B.~T.\ 2014a, MNRAS, 445, 2794

\bibitem[Veras et al.(2014b)]{veretal2014b} Veras D., Evans N.~W., Wyatt M.~C., Tout C.~A.\ 2014b, MNRAS, 437, 1127 

\bibitem[Veras, Shannon \& G\"{a}nscike(2014c)]{veretal2014cfor3} Veras D., Shannon A., G\"{a}nsicke B.~T.\ 2014c, MNRAS, 445, 4175

\bibitem[Veras, Eggl \& G\"{a}nscike(2015a)]{veretal2015afor3} Veras D., Eggl S., G\"{a}nsicke B.~T.\ 2015a, MNRAS, 451, 2814 

\bibitem[Veras, Eggl \& G\"{a}nsicke(2015b)]{veretal2015bfor3} Veras D., Eggl S., G\"{a}nsicke B.~T.\ 2015b, MNRAS, 452, 1945

\bibitem[Veras et al.(2015c)]{veretal2015c} Veras D., Brown D.~J.~A., Mustill A.~J., Pollacco D.\ 2015c, MNRAS, 453, 67 

\bibitem[Veras et al.(2016a)]{veretal2016a} Veras D., Mustill A.~J., G{\"a}nsicke B.~T., et al.\ 2016a, MNRAS, 458, 3942 

\bibitem[Veras et al.(2016b)]{veretal2016b} Veras D., Carter P.~J., Leinhardt Z.~M., G{\"a}nsicke B.~T., Submitted to MNRAS

  
\bibitem[Vial(2013)]{vial2013} Vial J.~C. 2013, Lecture Notes in Physics, Vol. 857. Berlin, Heidelberg: Springer.

\bibitem[Villaver \& Livio(2007)]{villiv2007} Villaver E., Livio M.\ 2007, ApJ, 661, 1192 

\bibitem[Villaver \& Livio(2009)]{villiv2009} Villaver E., Livio M.\ 2009, ApJL, 705, L81 

\bibitem[Villaver et al.(2014)]{viletal2014} Villaver E., Livio M., Mustill A.~J., Siess L.\ 2014, ApJ, 794, 3 

\bibitem[Voyatzis et al.(2013)]{voyetal2013} Voyatzis G., Hadjidemetriou J.~D., Veras D., Varvoglis H.\ 2013, MNRAS, 430, 3383 

\bibitem[Wilson et al.(2014)]{wiletal2014} Wilson D.~J., G{\"a}nsicke B.~T., Koester D., et al.\ 2014, MNRAS, 445, 1878 

\bibitem[Wyatt et al.(2014)]{wyaetal2014} Wyatt M.~C., Farihi J., Pringle J.~E., Bonsor A.\ 2014, MNRAS, 439, 3371 

\bibitem[Xu \& Jura(2014)]{xujur2014} Xu S., Jura M.\ 2014, ApJL, 792, L39 

\bibitem[Xu et al.(2016)]{xuetal2016} Xu S., Jura M., Dufour P., Zuckerman B.\ 2016, ApJL, 816, L22 

\bibitem[Zakamska \& Tremaine(2004)]{zaktre2004} Zakamska N.~L., Tremaine S.\ 2004, AJ, 128, 869 

\bibitem[Zeebe(2015)]{zeebe2015} Zeebe R.~E.\ 2015, ApJ, 798, 8 

\bibitem[Zhou et al.(2016)]{zhoetal2016} Zhou G., Kedziora-Chudczer L., Bailey J., et al.\ 2016, submitted to MNRAS, arXiv:1604.07405 

\bibitem[Zuckerman \& Becklin(1987)]{zucbec1987} Zuckerman B., Becklin E.~E.\ 1987, Nature, 330, 138 
  
\bibitem[Zuckerman et al.(2003)]{zucetal2003} Zuckerman B., Koester D., Reid I.~N., H\"{u}nsch, M.\ 2003, ApJ, 596, 477 

\bibitem[Zuckerman et al.(2010)]{zucetal2010} Zuckerman B., Melis C., Klein B., Koester D., Jura M.\ 2010, ApJ, 722, 725 

\end{thebibliography}
\end{document}